\documentclass[journal, twoside, final]{IEEEtran}
\usepackage{amsmath, amsfonts}
\usepackage{algorithmic}
\usepackage{algorithm}
\usepackage{array}
\usepackage[caption=false, font=scriptsize, labelfont=rm, textfont=rm]{subfig}
\usepackage{textcomp}
\usepackage{stfloats}
\usepackage{url}
\usepackage{verbatim}
\usepackage{graphicx}
\usepackage{cite}

\usepackage{amssymb, amsthm}
\usepackage{bm, color}
\usepackage{booktabs}
\usepackage{setspace}
\usepackage{url}
\usepackage{graphicx}
\usepackage{longtable}
\usepackage{makecell}

\usepackage{color}
\usepackage{dsfont}
\usepackage{datetime}
\usepackage{float}
\usepackage{bbm}
\usepackage{siunitx}
\usepackage{enumitem}
\usepackage{nicematrix}
\usepackage{multicol}

\newtheorem{assumption}{Assumption}
\newtheorem{definition}{Definition}

\newtheorem{lemma}{Lemma}

\newtheorem{theorem}{Theorem}

\allowdisplaybreaks[4]

\newcounter{mytempeqncnt}

\begin{document}
	\title{Air-to-Ground Communications Beyond 5G: UAV Swarm Formation Control and Tracking}
	
	\author{Xiao Fan, Peiran Wu, \IEEEmembership{Member,~IEEE}, and  Minghua Xia, \IEEEmembership{Senior~Member,~IEEE}
		\thanks{
			Manuscript received 02 January 2023; revised 13 August 2023 and 12 November 2023; accepted 14 December 2023. This work was supported in part by the National Natural Science Foundation of China under Grants U2001213 and 62171486 and in part by the Guangdong Basic and Applied Basic Research Project under Grant 2022A1515140166. {The material in this paper was partially presented as \cite{Fan2023} in the IEEE International Conference on Acoustic Speech Signal Process Workshop (ICASSPW), Rhodes Island, Greece, June 4--10, 2023.} The associate editor coordinating the review of this article and approving it for publication was H. Chen. {\it (Corresponding author: Minghua Xia.)}}
		\thanks{
			{Xiao Fan is with the School of Electronics and Information Technology, Sun Yat-sen University, Guangzhou 510006, China (e-mail: fanx26@mail2.sysu.edu.cn).} 
			
			{Peiran Wu and Minghua Xia are with the School of Electronics and Information Technology, Sun Yat-sen University, Guangzhou 510006, China, and also with the Southern Marine Science and Engineering Guangdong Laboratory, Zhuhai 519082, China (e-mail: \{wupr3, xiamingh\}@mail.sysu.edu.cn)).}}
		\thanks{
			{Color versions of one or more of the figures in this article are available online at https://ieeexplore.ieee.org.} 
			
			{Digital Object Identifier XXX}}
		}

	\markboth{IEEE Transactions on Wireless Communications}{Fan \MakeLowercase{\textit{et al.}}: Air-to-Ground Communications Beyond 5G}
	
	\maketitle

    \IEEEpubid{\begin{minipage}{\textwidth} \ \\[12pt] \centering 1536-1276 \copyright\ 2022 IEEE. Personal use is permitted, but republication/redistribution requires IEEE permission. \\
    See \url{https://www.ieee.org/publications/rights/index.html} for more information.\end{minipage}}

	\maketitle
	
	\begin{abstract}
		\noindent Unmanned aerial vehicle (UAV) communications have been widely accepted as promising technologies to support air-to-ground communications in the forthcoming sixth-generation (6G) wireless networks. This paper proposes a novel air-to-ground communication model consisting of aerial base stations served by UAVs and terrestrial user equipments (UEs) by integrating the technique of coordinated multi-point (CoMP) transmission with the theory of stochastic geometry. In particular, a CoMP set consisting of multiple UAVs is developed based on the theory of Poisson-Delaunay tetrahedralization. Effective UAV formation control and UAV swarm tracking schemes for two typical scenarios, including static and mobile UEs, are also developed using the multi-agent system theory to ensure that collaborative UAVs can efficiently reach target spatial positions for mission execution. Thanks to the ease of mathematical tractability, this model provides explicit performance expressions for a typical UE's coverage probability and achievable ergodic rate. Extensive simulation and numerical results corroborate that the proposed scheme outperforms UAV communications without CoMP transmission and obtains similar performance to the conventional CoMP scheme while avoiding search overhead.
	\end{abstract}
	
	
	 \IEEEpubidadjcol
	 	
	\section{Introduction} \label{sec: introduction}
	
	\IEEEPARstart{W}{ith} the rapid progress of unmanned aerial vehicles (UAVs) manufacturing technology and cost reduction, deploying UAVs in commercial applications has become popular in recent years, like Amazon's drone-delivery system. In the forthcoming sixth-generation (6G) wireless networks, aerial base stations served by UAVs are indispensable for three-dimensional (3D) air-to-ground networks \cite{Guo2022}. They can establish subnets independent of fixed networks, e.g., terrestrial and satellite networks, and provide a higher probability for line-of-sight (LoS) links by adjusting their altitudes and avoiding obstacles. Moreover, UAVs can be flexibly deployed cost-effectively to support ubiquitous coverage and high data-rate communications for intended areas, such as emergent disaster-relieves \cite{Ullah2020}. Despite appealing advantages, UAV communications also encounter many technical challenges, such as networking structure, movement control, transmission mechanisms, and performance characterization \cite{Mozaffari2019}. This paper will develop a mathematically tractable system model for air-to-ground communications and devise effective movement control for UAVs to effectively serve static and mobile terrestrial user equipments (UEs).

	 \IEEEpubidadjcol
	 	
	\subsection{Related Works and Motivation}
	Several recent works study air-to-ground wireless networks where UAVs serving as aerial base stations communicate with UEs. Given that all UAVs hover at the same fixed altitude, the effects of path loss exponent, height, and intensity of UAV were disclosed in \cite{Zhou2018}; and the approximate coverage probability for the 3D blockage effects was derived in \cite{Tang2020}. Given variable altitudes of UAVs, the coverage and network throughput of non-orthogonal multiple access (NOMA)-assisted UAV network modeled by a 3D homogeneous Poisson point process (PPP) in the spherical space were analyzed in \cite{Hou2020}, while the work \cite{Liu2021} investigated the coverage performance of UAV-enabled cellular networks modeled by a 3D PPP. 
		 	
	Coordinated multi-point (CoMP) techniques, a key technique in LTE-Advanced \cite{3GPPTR36.819}, have also been applied to enhance UAV-enabled wireless communications. In principle, CoMP techniques exploit more than one BS forming a cooperation set and simultaneously serving a user to improve quality of service (QoS), especially for users at the cell edge \cite{9644611}. A user-centric cooperative UAV clustering scheme was proposed in \cite{Wu2018}, and the work \cite{Liu2019} designed the optimal trajectories of multiple UAVs that maximize the uplink throughput according to real-time locations of UEs. The QoS of UAV-assisted cellular networks was optimized in \cite{Zhu2020} by jointly designing access point selection and UAV path planning. In contrast, the NOMA scheme was exploited for cellular-connected UAVs in \cite{New2022} to ensure high spectral efficiency and massive connectivity for both aerial and terrestrial UEs. 
	
	Unlike the above works considering UAV-assisted cellular networks, the scenarios only with the UAVs, which are an essential part of the non-terrestrial network, have gained attention in recent years. The work \cite{Qiu2020} studied the joint optimization of resource allocation, UAV placement, and user association to improve the throughput of UEs within the flight time of UAVs. The optimal 3D locations and the minimum number of UAVs were investigated in \cite{Zhang2021} to meet the target coverage and band allocation requirements. In practice, those approaches are achieved at the cost of an overwhelming searching complexity and feedback overhead. In contrast, the triangulation theory can serve a network architecture called the Delaunay simplex\footnote{For illustration purposes, we consider a large-scale cellular system where the locations of the BSs are modeled as a PPP. Suppose all BSs transmit with the same power and each user associates with one BS as per the nearest neighbor criterion, the coverage regions of the BSs form a Poisson-Voronoi tessellation, and its dual graphs are called the Poisson-Delaunay simplex. The typical cell of Poisson-Voronoi tessellation is an irregular polygon with an uncertain number of edges, and some basic features of the typical cell are still unknown. In contrast, the Poisson-Delaunay simplex has regular cells. For instance, the typical cell of Poisson-Delaunay simplex in 3D is a tetrahedron, yielding Poisson-Delaunay tetrahedralization. For more details on Delaunay simplex, the interested reader is referred to \cite{Hjelle2006}.}, which can be fixed and uniquely determined by the geometric locations of its nearby UAVs \cite{Xia2018}. This simplex was applied in recent works \cite{Li2020A2A, Li2022A} to establish the CoMP transmission set of air-to-air and ground-to-air networks, respectively. 
	
	To study the mobility of UAVs forming the required Delaunay simplex, the movement control of UAVs is imperative. In the open literature, movement control has been studied primarily from a robotics/control perspective. More specifically, movement control aims to design an appropriate control strategy to drive agents to form the required geometric pattern and guide them to maneuver as a whole \cite{Shakeri2019}. In this regard, the integrated controller was designed in \cite{Zhang2021june}, which can control the velocity consensus for the UAVs' formation and keep the formation as a fixed geometry. The work \cite{Qian2021} jointly analyzed the stability of formation control and the capacity of wireless communications between agents moving in a formation.
	
	According to fundamental ideas in control schemes, formation control consists of three strategies: leader-follower, behavioral, and virtual structure \cite{Oh2015}. Among them, the leader-follower strategy is the most common since it can design simple or complex formation tracking controllers. The work \cite{Wu2020} addressed a formation tracking control problem for a second-order nonlinear multi-agent system with multiple leaders via distributed impulsive control methods. The study \cite{Ringback2021} designed a distance-based formation control algorithm with a collision-avoidance potential function to solve a collaborative tracking problem. However, the trackers in these works mentioned above had the same dynamics as the target, which is not always accurate in practice. Also, the idea of the practical consensus \cite{Bernuau2019}, which is rarely used in UAV formation control, can be exploited to save the limited power supply of UAVs.
	
	Motivated by the above considerations and particularly inspired by the recent work \cite{Li2020A2A}, the binomial-Delaunay tetrahedralization is exploited in this paper to model 3D height-limited spatial deployment of UAVs, which are indispensable for future space-air-ground-sea integrated networks \cite{Geraci2022}. With this geometric method, a 3D air-to-ground CoMP transmission scheme is proposed. Then, by exploiting the Lyapunov function method in stability theory \cite{Cao2013}, the UAV formation control for static UEs and UAV swarm tracking for mobile UEs are designed, followed by analytical performance analysis.
	
	\subsection{Summary of Major Contributions}
	In this paper, we first develop a mathematically tractable UAV-enabled wireless communication architecture based on CoMP transmission. Then, the practical movement control strategies for two cases, including UAV formation control for static UEs and UAV swarm tracking for mobile UEs, are designed. Afterward, the coverage probability and ergodic rate for a typical UE are analytically derived. In a nutshell, the significant contributions of this paper include: 
	\begin{enumerate} [label = {\arabic*)}]
		\item Network model: A novel 3D UAV-enabled wireless network model is devised based on the theory of Poisson-Delaunay tetrahedralization. In particular, a CoMP strategy is designed for the joint transmission of four UAVs to improve communication quality. 
		\item UAV formation control for {\it static} UEs: To ensure the QoS of static UEs, the UAV formation during the process of takeoff and flight, i.e., the geometric pattern of UAVs determined by the CoMP transmission scheme is controlled by a pining control strategy.
		\item UAV swarm tracking control for {\it mobile} UEs: To track a mobile UE, the UAV swarm with a given geometric pattern is controlled by an impulsive control strategy, which allows a piece-wise adaptive control even if mobile UEs have different dynamics from UAVs.
		\item Analytical performance analysis: The coverage probability and ergodic rate are explicitly derived for a typical UE. Simulation and numerical results corroborate that the proposed scheme outperforms UAV communications without CoMP transmission and obtains similar performance to the conventional CoMP scheme while avoiding search overhead.
	\end{enumerate}

	\subsection{Paper Organization}
	The remainder of this paper is organized as follows. Section~\ref{sec: model} proposes a novel air-to-ground communication system model. Section~\ref{sec: control1} develops a UAV formation control strategy for static UEs, while Section~\ref{sec: control2} devises a UAV swarm tracking strategy for mobile UEs. Section~\ref{sec: performance}  analyzes the performance of a typical UE in terms of the coverage probability and ergodic rate. Simulation and numerical results are presented and discussed in Section~\ref{sec: simulation}. Finally, Section~\ref{sec: conclusion} concludes the paper.
	
	{\it Notation:} Scalars, vectors, and matrices are denoted by italic letters, lower- and uppercase letters in bold typeface, respectively. The symbols $\mathbb{R}^{n}$, $\mathbb{R}_{+}$, and $\mathbb{N}_{+}$ indicate the real space of dimension $n$, the set of positive real numbers, and the set of natural numbers, respectively. The symbols $|\cdot|$, $\|\cdot\|$, and $\|\cdot\|_{\infty}$ denote the modulus of a complex-valued number, $\ell_2$-norm, and $\ell_{\infty}$-norm of vector/matrix, respectively. The superscripts $(\cdot)^{\text{T}}$ and $(\cdot)^{\text{H}}$ represent the transpose and conjugate transpose, respectively. The symbols $\bm{0}$, $\mathbbm 1$, $\bm{O}$, and $\bm{I}$ denote all-zero vector, all-ones vector, all-zero matrix, and identity matrix of appropriate size, respectively; ${\rm Diag}\{x_1, \cdots, x_n\}$ denotes a diagonal matrix with diagonal entries $x_1, \cdots, x_n$; and $\otimes$ denotes the Kronecker product. The symbols $\lambda_{\min} (\cdot)$ and $\lambda_{\max} (\cdot)$ stand for the minimum and maximum eigenvalues of a matrix, respectively. The probability density function (PDF) and complementary cumulative distribution function (CCDF) of a random variable $X$ are denoted by $f_{x}(\cdot)$ and $F_{x}^{c}(\cdot)$, respectively. The operators $\mathbb{E} [\cdot]$ and ${\text{Var}} [\cdot]$ compute the expectation and variance of a random variable, respectively. The symbol ${\binom{n}{m}} \triangleq \frac{n !}{m ! (n - m) !}$ refers to the binomial coefficient, with $n !$ being the factorial of a positive integer $n$. The Gamma and upper incomplete Gamma functions are defined as $\Gamma(a) \triangleq \int_{0}^{\infty} t^{a - 1} \exp(- t) \,  {\rm d} t$ and $\Gamma(a, x) \triangleq \int_{x}^{\infty} t^{a - 1} \exp(- t) \, {\rm d} t$, respectively. The function $P_{a}^{(n)}(x) \triangleq \sum_{k = 0}^{n} (-1)^{k} {\binom{n}{k}} \Gamma(k + 1 - a) x^{k}$ is an $n^{\rm th}$-order polynomial of $x$ with parameter $a$. The parabolic cylinder function is defined as $D_{p}(z) \triangleq 2^{\frac{p}{2}} \exp\left(-z^{2}/4\right) \left[\frac{\sqrt{\pi}}{\Gamma\left((1 - p)/2\right)} {_{1}F_{1}} \left(-\frac{p}{2}; \frac{1}{2}; \frac{z^{2}}{2}\right) \right.$ $\left.- \frac{\sqrt{2 \pi} z}{\Gamma\left(- p/2 \right)} {_{1}F_{1}} \left(\frac{1 - p}{2}; \frac{3}{2}; \frac{z^{2}}{2}\right)\right],$ with degenerate hypergeometric function ${_{1}F_{1}} (a; b; z) \triangleq \sum_{n = 0}^{\infty} \frac{\Gamma(a + n) \Gamma(b)}{\Gamma(a) \Gamma(b + n)} \frac{z^{n}}{n !}$. These special functions can be computed using built-in functions in regular numerical software, such as MATLAB and Mathematica.

	\section{System Modeling of UAV-enabled Wireless Networks} \label{sec: model}
\begin{figure}[!t]
    	\centering
    	\includegraphics[width=0.8\linewidth]{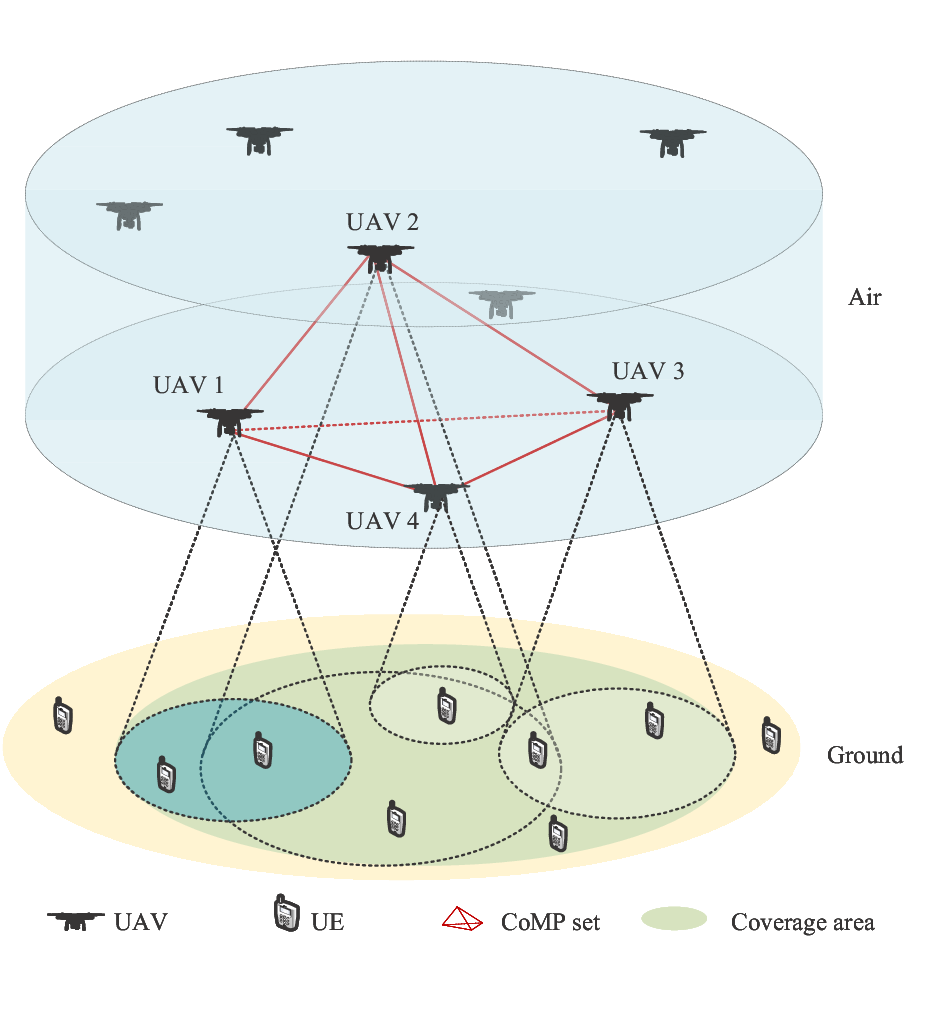}
    	\vspace{-10pt}
    	\caption{System model of a UAV-enabled wireless network, where UAVs are distributed in a 3D height-limited space.}
    	\label{Fig: model}
    \end{figure}

	As illustrated in Fig.~\ref{Fig: model}, we consider the downlink transmission of a UAV-enabled wireless network, where UAVs acting as aerial base stations are deployed in a 3D height-limited space to provide LoS data transmission to terrestrial UEs. Conventionally, one UAV can serve multiple terrestrial UEs in a relatively small coverage area (e.g., the small blue area covered by UAV~1 in Fig.~\ref{Fig: model}). To serve more UEs with improved QoS, several UAVs can collaborate and serve a broader coverage (e.g., the large green area jointly covered by UAVs 1-4 in Fig.~\ref{Fig: model}). In this regard, the CoMP technique can be exploited to enable the joint transmission of multiple UAVs. In theory, more collaborative UAVs yield better QoS but with higher communication overhead. In practice, determining the best cooperative UAVs is usually based on exhaustive searching, e.g., using the nearest neighbor criterion\cite{Feng2019}. This traditional method is time-consuming and needs channel state information (CSI) estimation, which is hard to implement due to the fast speed of UAVs. Also, previous studies, e.g., \cite{Li2020A2A, 9644611}, show that four cooperative entities can almost obtain the largest cooperation gain. Consequently, in the following, we exploit the theory of Poisson-Delaunay tetrahedralization to determine the CoMP set of UAVs.
    
	Like \cite{Liu2021}, the UAVs are supposed to be distributed according to a marked PPP, i.e.,
	\begin{equation} \label{Eq: marked-PPP}
		\Phi_{B} \triangleq \left\{\bm{B}_{i} = \left(\bm{x}_{i}, h_{i}\right) \in \mathbb{R}^{2} \times \mathbb{R}_{+}: \bm{x}_{i} \in \Phi_{x}, i \in \mathbb{N}_{+}\right\},
	\end{equation}
	where $\bm{x}_{i}$ and $h_{i}$ denote the planar locations and height of the $i^{\rm th}$ UAV, respectively, and they are independent with each other; $\Phi_{x} \in \mathbb{R}^{2}$ is a homogeneous PPP with density $\lambda_{x}$, which specifies the geographic locations $\bm{x}_{i}$ of the planar projection of UAVs, $\forall i \in \mathbb{N}_{+}$; and the height $h_{i} \in [H_{\min},  H_{\max}]$, with $H_{\min}$ and $H_{\max}$ being the minimum and maximum height of UAVs, respectively. On the other hand, the terrestrial UEs are distributed as per another homogeneous PPP $\Phi_{U} \in \mathbb{R}^{2}$, which has density $\lambda_U$ and is independent of $\Phi_{x}$. 
	
\begin{figure}[!t]
    	\centering
    	\includegraphics[width=0.98\linewidth]{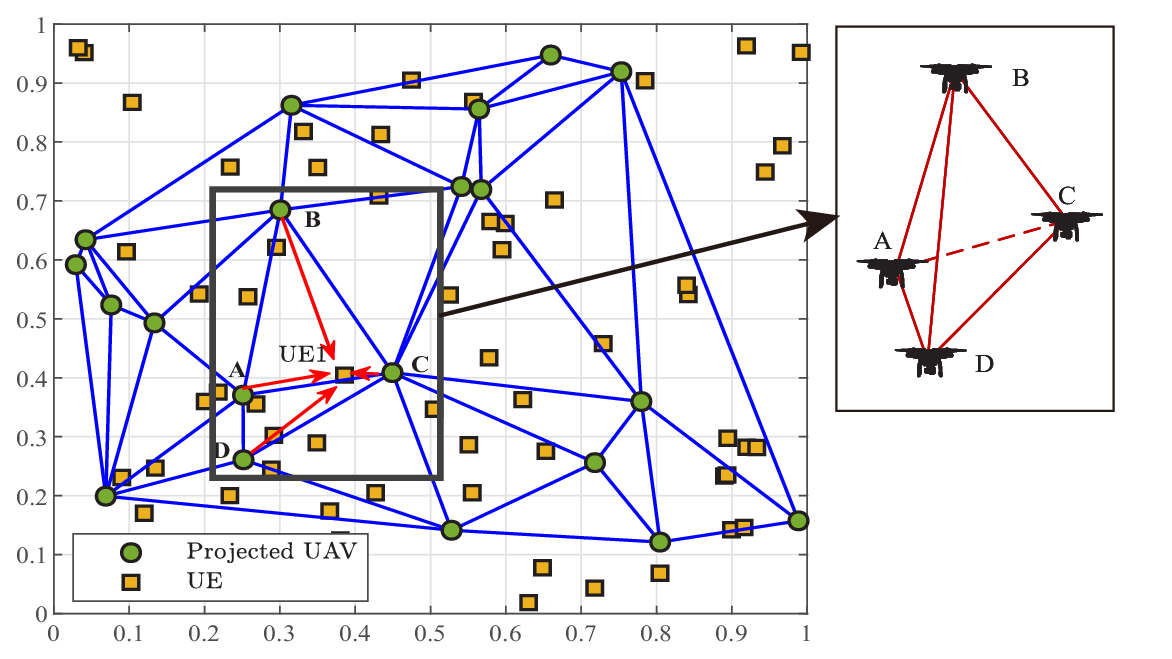}
    	\caption{An illustrative 2D Poisson-Delaunay triangulation network where the projected UAVs (the circles) and UEs (the squares) are distributed as per their respective PPPs.  The CoMP sets are determined by the Delaunay triangles (solid blue boundaries). The right subfigure shows the spatial positions of the collaborative UAVs.}
    	\label{Fig: compset}
    \end{figure}

	Based on the planar projection of UAVs, the 2D Poisson-Delaunay simplex is uniquely determined and can be constructed by efficient algorithms in the triangulation theory, such as the divide-and-conquer algorithm \cite[Ch. 4]{Hjelle2006}. Then, to construct Delaunay tetrahedralization in 3D space, we consider the cooperating set $\Phi_{0}$, composed of the four nearest neighboring UAVs of a typical UE. As the UAVs are located in a height-limited space, the Euclidean distance between the UAV and UE is mainly determined by the horizontal distance between them. Then $\Phi_{0}$ can be approximately determined by the set of four ``nearest'' neighboring UAVs of a typical UE, where the ``nearest'' is in the sense of horizontal distance between the location of the planar projection of a UAV and a typical UE under study. For illustration purposes, Fig.~\ref{Fig: compset} shows that a typical UE~$1$ first chooses the nearest UAV $C$ and the second-nearest UAV $A$. We follow this process until the CoMP set of UE~$1$ consisting of four UAVs $\{A, B, C, D\}$ is formed. 
	
	To shape the geometric pattern related to our CoMP transmission scheme, a collaborative UAV formation control strategy for {\it static} UEs needs to be designed by accounting for UAVs' maneuverability. On the other hand, to track {\it mobile} UEs, it is essential to develop an energy-efficient tracking strategy for a UAV swarm since the mechanical energy consumption of UAVs dominates over electrical energy consumption. These two strategies will be elaborated on each in the following two sections. 
	
	\section{UAV Formation Control for Static UEs} \label{sec: control1}
	To serve static terrestrial UEs, we must steer UAVs from the take-off points to reach the target spatial positions and form the geometric pattern our CoMP scheme determines. Accordingly, by taking the UAVs as agents in the network, we exploit graph theory to theoretically model UAVs' communication and nonlinear multi-agent system theory to analyze the dynamics of a UAV swarm.
	
	\subsection{The Dynamics of a UAV Swarm} \label{sec: control1-a}
	We model a UAV swarm as a nonlinear multi-agent system consisting of one leader and $n$ followers. The interaction among the followers can be described as a weighted digraph $\mathcal{G} = (\mathcal{V}, \mathcal{E})$, with the set of nodes $\mathcal{V} = \left\{1, \cdots, n \right\}$ and the set of directed edges $\mathcal{E} \subseteq \mathcal{V} \times \mathcal{V}$. A directed edge $a_{i j}$ in $\mathcal{E}$ is denoted by the ordered pair of nodes $\left(i, j\right)$, which means node $j$ can get information from node $i$, yet not vice versa. Define the weighted adjacency matrix ${\bm A} = \left[a_{i j}\right] \in \mathbb{R}^{n \times n}$ with $a_{i j} = 0$ if $i = j$ and $a_{i j} > 0$ if $i \neq j$, and ${\bm D} = \left[d_{i j}\right] \in \mathbb{R}^{n \times n}$ being the in-degree matrix with $d_{i j} = \sum_{j = 1}^{n} a_{i j}$ if $i = j$ and $d_{i j} = 0$ if $i \neq j$. Finally, let $\bm{L}  \triangleq {\bm D} - {\bm A} \in \mathbb{R}^{n \times n}$ be the Laplacian matrix associated with $\mathcal{G}$. 
	
	In the following, we use a point above a function to mean its first-order differential. According to \cite{Wang2017}, the dynamics of the multi-agent system formulated above can be described as: \vspace{-15pt}
	\begin{subequations} \label{Eq: C1-general} 
		\begin{align}
			\dot{\bm{x}}_{0} (t) & = {\bm{v}}_{0} (t), \label{Eq: C1-leader-a} \\
			\dot{\bm{v}}_{0} (t) & = \bm{f} \left(t, \bm{x}_{0} (t), \bm{v}_{0} (t)\right), \label{Eq: C1-leader-b} \\
			\dot{\bm{x}}_{i} (t) & = \bm{v}_{i} (t), \label{Eq: C1-follower-a} \\
			\dot{\bm{v}}_{i} (t) & = {\bm f} \left(t, \bm{x}_{i} (t), \bm{v}_{i} (t)\right) + \sum_{j = 1}^{n} a_{i j} \left[c_{x} \left(\bm{x}_{j} (t) - {\bm{x}}^{\ast}_{j}  \right.\right. \nonumber \\
			& \quad \left. \left. - \bm{x}_{i} (t) + {\bm{x}}^{\ast}_{i}\right) + c_{v} \left(\bm{v}_{j} (t) - \bm{v}_{i} (t)\right)\right]  + {\bm u}_{i} (t), \label{Eq: C1-follower-b}
		\end{align} 
	\end{subequations}
	for $i = 1, \cdots, n$, where the vector $\bm{x}_{0} (t) \in \mathbb{R}^{m}$ in \eqref{Eq: C1-leader-a} denotes the position of the leader, and the first-order differential $\dot{\bm{x}}_{0} (t)$ with respect to $t$ means its velocity given by $\bm{v}_{0} (t)$ in \eqref{Eq: C1-leader-a}; the function ${\bm f} \left(\cdot\right)$ in \eqref{Eq: C1-leader-b} represents the intrinsic nonlinear dynamics of the agents in which the location information of UE is implicitly involved; $\bm{x}_{i} (t)$, $\bm{v}_{i} (t)$, and ${\bm u}_{i} (t) \in \mathbb{R}^{m}$ in  \eqref{Eq: C1-follower-a}-\eqref{Eq: C1-follower-b} denote the position, velocity, and control vector of the $i^{\rm th}$ follower at time $t$, respectively;  $\bm{x}_{i}^{\ast}$ in \eqref{Eq: C1-follower-b}  represents the expected formation position of the $i^{\rm th}$ follower and finally, $c_{x}$ and $c_{v}$ in \eqref{Eq: C1-follower-b} are control constants and they usually take the same values, i.e., $c_{x} = c_{v} \triangleq c$ \cite{Wang2017}. For the sake of notational simplicity, all independent variables except $t$ in the function $\bm{f}(\cdot)$ are omitted if no confusion arises. For instance, $\bm{f} \left(t, \bm{x}_{0} (t), \bm{v}_{0} (t)\right)$ and ${\bm f} \left(t, \bm{x}_{i} (t), \bm{v}_{i} (t)\right)$ are abbreviated as $\bm{f}_{0} (t)$ and $\bm{f}_{i}(t)$, respectively.
	
	\subsection{UAV Swarm Movement Control} \label{sec: control1-b}	
	It is obvious that \eqref{Eq: C1-general} is a second-order system. Before moving forward, we introduce a criterion on the second-order leader-follower flight formation control.
	
	\begin{definition}[Def. 5 of {\cite{Yu2010}}]
		\label{Def: formation}
		A multi-agent system achieves second-order leader-follower flight formation control if, for any initial state of agent $i$, the solution to \eqref{Eq: C1-general} satisfies
		\begin{subequations} \label{Eq-Def1}
			\begin{align}
				\lim_{t \to \infty} \left\|\bm{x}_{i} (t) - \bm{x}_{0} (t) - {\bm{x}}^{\ast}_{i}\right\| &= 0, \label{Eq-Def1-a} \\
				\lim_{t \to \infty} \left\|\bm{v}_{i} (t) - \bm{v}_{0} (t)\right\| &= 0. \label{Eq-Def1-b}
			\end{align}
		\end{subequations}	
	\end{definition}
	
	Now, we adopt a pinning control strategy to solve \eqref{Eq: C1-general}. Accordingly, the control vector in \eqref{Eq: C1-follower-b} is determined as \cite{Oh2015} 
	\begin{equation}
		\bm{u}_{i} (t) = - c \, b_{i} \left[\left(\bm{x}_{i} (t) - \bm{x}_{0} (t) - {\bm{x}}^{\ast}_{i}\right) + \left(\bm{v}_{i} (t) - \bm{v}_{0} (t)\right)\right],
		\label{Eq: C1-pining control}
	\end{equation}
	where $b_{i} \geq 0$ denotes the local feedback gain obtained at the $i^{\rm th}$ agent. Then, the position and velocity errors can be computed as
	\begin{subequations} \label{Eq: C1-error_0}
		\begin{align}
			\bm{\xi}_{\bm{x}} (t) & = \left[\begin{smallmatrix} (\bm{x}_{1} (t) - \bm{x}_{0} (t) - \bm{x}_{1}^{\ast})^{\text T}, & \cdots, & (\bm{x}_{n} (t) - \bm{x}_{0} - \bm{x}_{n}^{\ast})^{\text T} \end{smallmatrix}\right]^{\text T},
			\label{Eq: C1-error-x} \\
			\bm{\xi}_{\bm{v}} (t) & = \left[\begin{smallmatrix}  (\bm{v}_{1} (t) - \bm{v}_{0} (t))^{\text T}, & \cdots, & (\bm{v}_{n} (t) - \bm{v}_{0})^{\text T} \end{smallmatrix}\right]^{\text T}.
			\label{Eq: C1-error-v} 
		\end{align}
		Combining \eqref{Eq: C1-error-x} and \eqref{Eq: C1-error-v}, we define  
		\begin{equation}
			\bm{\epsilon} (t) \triangleq \begin{bmatrix}\bm{\xi}_{\bm{x}}^{\text T} (t), & \bm{\xi}_{\bm{v}}^{\text T} (t) \end{bmatrix}^{\text T}.
			\label{Eq: C1-error-xv}
		\end{equation}
	\end{subequations}
	In light of \eqref{Eq: C1-general}, \eqref{Eq: C1-pining control} and \eqref{Eq: C1-error_0}, taking the first derivative of \eqref{Eq: C1-error-xv} with respect to $t$ and after making some algebraic manipulations, we get the error dynamics expressed as
	\begin{equation}\label{Eq: C1-error}
		\dot{\bm{\epsilon}} (t)  =
		\left[\begin{smallmatrix}
			\bm{O}_{n} 							& {\bm I}_{n} \\
			- c \left(\bm{L}  + \bm{B} \right)	& - c \left(\bm{L}  + \bm{B} \right)
		\end{smallmatrix}\right]
		\otimes \bm{I} _{m} {\bm \epsilon} (t) +
		\left[\begin{smallmatrix}
			\bm{0}_{mn} \\
			{\bm F} \left(t\right) - \mathbbm{1}_{n} \otimes {\bm f}_{0} \left(t\right)
		\end{smallmatrix}\right],
	\end{equation}
	where $\bm{F} (t) = [{\bm f}_{1}^{\text T} (t), \cdots,  {\bm f}_{n}^{\text T}(t) ]^{\text T}$, and $\bm{B} \in \mathbb{R}^{n \times n}$ is a diagonal matrix with the $i^{\rm th}$ diagonal entry specified by $b_{i}$. 
	
	Before we present the main result of this section, two preliminary assumptions in stability theory are reproduced below.
	\begin{assumption}[Strongly connected]
		\label{Asump: graph}
		The graph $\mathcal{G}$ regarding the system given by \eqref{Eq: C1-general} is strongly connected, i.e., there is a path in each direction between each pair of vertices of the graph.
	\end{assumption}
	
	It is evident that flight formation control cannot be achieved if an isolated node does not receive information from other nodes. As a result, this strong connection assumption certainly holds for the UAV swarm under study.		
	\begin{assumption}[Lipschitz condition]
		\label{Asump: Lipschitz}
		There exist two non-negative constants $\rho_{1}$ and $\rho_{2}$ such that
		\begin{align}
			\left\|\bm{f}_{i} (t) - \bm{f}_{j} (t)\right\| \leq \rho_{1} \left\|\bm{x}_{i} (t) - \bm{x}_{j} (t)\right\| + \rho_{2} \left\|\bm{v}_{i} (t) - \bm{v}_{j} (t)\right\|,
		\end{align}
		for all $\bm{x}_{i} (t)$, $\bm{v}_{i} (t)$, $\bm{x}_{j} (t)$, and $\bm{v}_{j} (t) \in \mathbb{R}^{m}$.
	\end{assumption}
	
	Based on the physical characteristics of agents, almost all intrinsic dynamics are consistent with Lipschitz condition \cite{Yu2010}. Thus, it is assumed feasible for the UAV swarm under study. 
	
	We are in a position to show the main result of this section. Given the initial states of agents:
	\begin{subequations} \label{Eq: C1-initial}
		\begin{align}
			\bm{x}(0) & = \begin{bmatrix} \bm{x}_{0}^{\text{T}} (0), \bm{x}_{1}^{\text{T}} (0), \cdots, \bm{x}_{n}^{\text{T}} (0) \end{bmatrix} ^{\text{T}}, \\
			\bm{v}(0) & = \begin{bmatrix} \bm{v}_{0}^{\text{T}} (0), \bm{v}_{1}^{\text{T}} (0), \cdots, \bm{v}_{n}^{\text{T}} (0) \end{bmatrix} ^{\text{T}},
		\end{align}
	\end{subequations}
	the achievability of the flight formation control of a multi-agent system can be determined according to the following theorem.
	
	\begin{theorem} \label{Thm: formation}
		Given Assumptions \ref{Asump: graph} and \ref{Asump: Lipschitz}, for any initial state \eqref{Eq: C1-initial}, the second-order leader-follower flight formation control of the multi-agent system~\eqref{Eq: C1-general} with the control strategy \eqref{Eq: C1-pining control} is achievable if the control constant $c_{x} = c_{v} \triangleq c$ in \eqref{Eq: C1-follower-b} satisfies
		\begin{equation} \label{Eq: thm}
			c > \frac{d \, \lambda_{\max} \left({\bm P}\right)}{\lambda_{\min} \left({\bm Q}\right)},
		\end{equation}
		where $d = \max \left\{3 \rho_{1} + \rho_{2}, \rho_{1} + 3 \rho_{2} + 2\right\}$, $\bm{P} = {\rm Diag} \left\{1 / q_{1}, \cdots, 1 / q_{n}\right\}$ with $\bm{q} \triangleq \left[q_{1}, \cdots, q_{n}\right]^{\rm T} = \left(\bm{L}  + \bm{B} \right)^{- 1} \mathbbm{1}_{n}$, and $\bm{Q} = \bm{P} \left(\bm{L}  + \bm{B} \right) + \left(\bm{L}  + \bm{B} \right)^{\rm T} \bm{P}$.
	\end{theorem}
	
	\begin{IEEEproof}
		See Appendix \ref{Appendix: C1}.
	\end{IEEEproof}
	
	As Theorem~\ref{Thm: formation} guarantees the global asymptotic stability of the multi-agent system, the control constant $c$ is not necessarily set to global identity. In real-world applications, only the condition~\eqref{Eq: thm} needs to be satisfied to balance the flight speed and adequate control time of UAVs.
	
	It is noteworthy that environmental error is not accounted for in the dynamics of the multi-agent system given by \eqref{Eq: C1-general}. To be more practical, we consider a multi-agent system with additive noise. In this case, with control strategy \eqref{Eq: C1-pining control}, the term $c \left(\bm{L} + \bm{B} \right)$ in error dynamics \eqref{Eq: C1-error} can be replaced by $c \left(\bm{L} + \bm{B} \right) - \bm{E}$, where the error matrix ${\bm E} = (e_{i j}) \in \mathbb{R}^{n \times n}$ denotes the additive white noise whose power satisfies a standard Brownian motion. Then, with the same condition~\eqref{Eq: thm} in Theorem~\ref{Thm: formation}, the multi-agent system with additive noise can also achieve the second-order leader-follower flight formation control in the sense of mean squared error. That is, for any initial state, unlike \eqref{Eq-Def1-a}-\eqref{Eq-Def1-b}, the solution to the system with additive noise satisfies 
	\begin{subequations} \label{Eq: AWGN}
		\begin{align}
			\lim_{t \rightarrow + \infty} \mathbb{E} \left[\left\|\bm{x}_{i} (t) - \bm{x}_{0} (t) - {\bm{x}}^{\ast}_{i}\right\|^{2}\right] & = 0,   \label{Eq: AWGN-a} \\
			\lim_{t \rightarrow + \infty} \mathbb{E} \left[\left\|\bm{v}_{i} (t) - \bm{v}_{0} (t)\right\|^{2} \right] & = 0.  \label{Eq: AWGN-b}
		\end{align}
	\end{subequations}
	The proof is similar to that of Theorem~\ref{Thm: formation} and omitted here for brevity.
	
	\subsection{Case Study}  \label{sec: control1-c}	
    \begin{figure}[t]
    	\centering
    	\includegraphics[width=0.8\linewidth]{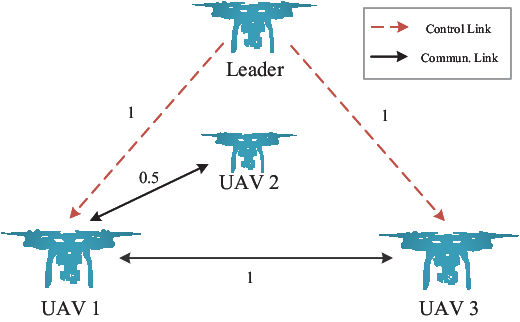}
    	\caption{An illustrative model for the communications between UAVs.}
    	\label{Fig: case1-graph}
    \end{figure}
	
	To illustrate the effectiveness of the control strategy \eqref{Eq: C1-pining control} that enables the geometric pattern related to our CoMP transmission scheme, Fig.~\ref{Fig: case1-graph} shows a multi-agent system consisting of four UAVs, where one of them serves as the leader and the others as the followers. It is clear that the system dynamics can be described by \eqref{Eq: C1-general} with $n = 3$, where the position and velocity states of the $i^{\rm th}$ UAV are denoted by $\bm{x}_{i}(t)$ and $\bm{v}_{i}(t) \in \mathbb{R}^{3}$, for all $i\in \{ 0, 1, 2, 3 \}$, respectively. Suppose the flight speed of each UAV is up to $20$ \si{m/s} and the UAV swarm is expected to reach above the origin $\bm{o} = (0, 0, 0)$ at the specified altitude $150$~\si{m} after formation.\footnote{According to the technical report 3GPP TR 23.754 \cite{3GPPTR23.754},  the flight height of UAV is limited between $50$~\si{m} and $300$~\si{m}. Accordingly, the leader's altitude is set to $150$~\si{m} in the pertaining simulation experiments.} Then, the nonlinear function $\bm{f}_{i} (t)$ used in \eqref{Eq: C1-follower-b} can be given by 
	\begin{equation}\label{Eq: case1-f}
		\bm{f}_{i} (t) = 0.0001
		\begin{bmatrix}
			1, &
			\dfrac{1}{{x_{i Y}^{2} + 1}}, &
			x_{i Z} - 150
		\end{bmatrix}^{\rm T},
	\end{equation}
	where the factor $0.0001$ aims to keep the speed within practical limits, and the three dimensions of position vectors $\bm{x}_{i}(t) = [x_{i X}(t), x_{i Y}(t), x_{i Z}(t)]^{\rm T}$, for all $i\in \{ 0, 1, 2, 3 \}$, correspond to the three axes $X$, $Y$, and $Z$ of the inertial coordinate system, and they follow the right-hand rule.
	
	According to Fig.~\ref{Fig: case1-graph}, the interconnection configuration for followers is a chain structure $\mathcal{G}$, whose in-degree matrix $\bm{D}$ and Laplacian matrix $\bm{L}$ are easily expressed as 
	\begin{equation*}
		\bm{D} =
		\begin{bNiceMatrix}[first-row, last-col, code-for-first-row=\scriptstyle, code-for-last-col=\scriptstyle, code-for-first-row = \color{black}, code-for-last-col = \color{black}]
			{\text {\tiny UAV 1}} & {\text {\tiny UAV 2}} & {\text {\tiny UAV 3}} &  \\
			1.5 & 0    & 0 & {\text {\tiny UAV 1}}  \\
			0   &  0.5 & 0 & {\text {\tiny UAV 2}} \\
			0   &  0   & 1 & {\text {\tiny UAV 3}}
		\end{bNiceMatrix}, \quad
		\bm{L} =
		\begin{bNiceMatrix}[first-row, last-col, code-for-first-row=\scriptstyle,
			code-for-first-row = \color{black}, code-for-last-col=\scriptstyle, code-for-last-col = \color{black}]
			{\text {\tiny UAV 1}} & {\text {\tiny UAV 2}} & {\text {\tiny UAV 3}} &  \\
			1.5  & -0.5    & -1 & {\text {\tiny UAV 1}}  \\
			-0.5 &  0.5 & 0 & {\text {\tiny UAV 2}} \\
			-1  &  0   & 1 & {\text {\tiny UAV 3}}
		\end{bNiceMatrix}.
	\end{equation*}
	It is noteworthy that the connection weight between followers directly controlled by the leader is higher than that between followers indirectly controlled, e.g., $a_{13} = 1$ while $a_{12} = 0.5$ in Fig.~\ref{Fig: case1-graph}. Also, the communication between the leader and followers in Fig.~\ref{Fig: case1-graph} can be reflected by $\bm{B} = {\text{Diag}} \left\{1, 0, 1\right\}$, which means the leader only controls UAVs~1 and 3, regardless of UAV~2. Furthermore, the coordinates of the ideal formation location $\bm{x}^{\ast}$ involved in \eqref{Eq: C1-follower-b} are obtained from a randomly chosen tetrahedral cell's coordinate from the Poisson-Delaunay tetrahedralization.
	
    \begin{figure*}
    	\captionsetup[subfigure]{margin=5pt}
    	\subfloat[The position and velocity errors]{
    		\label{Fig: case1-xv}
    		\includegraphics[width=0.7\linewidth]{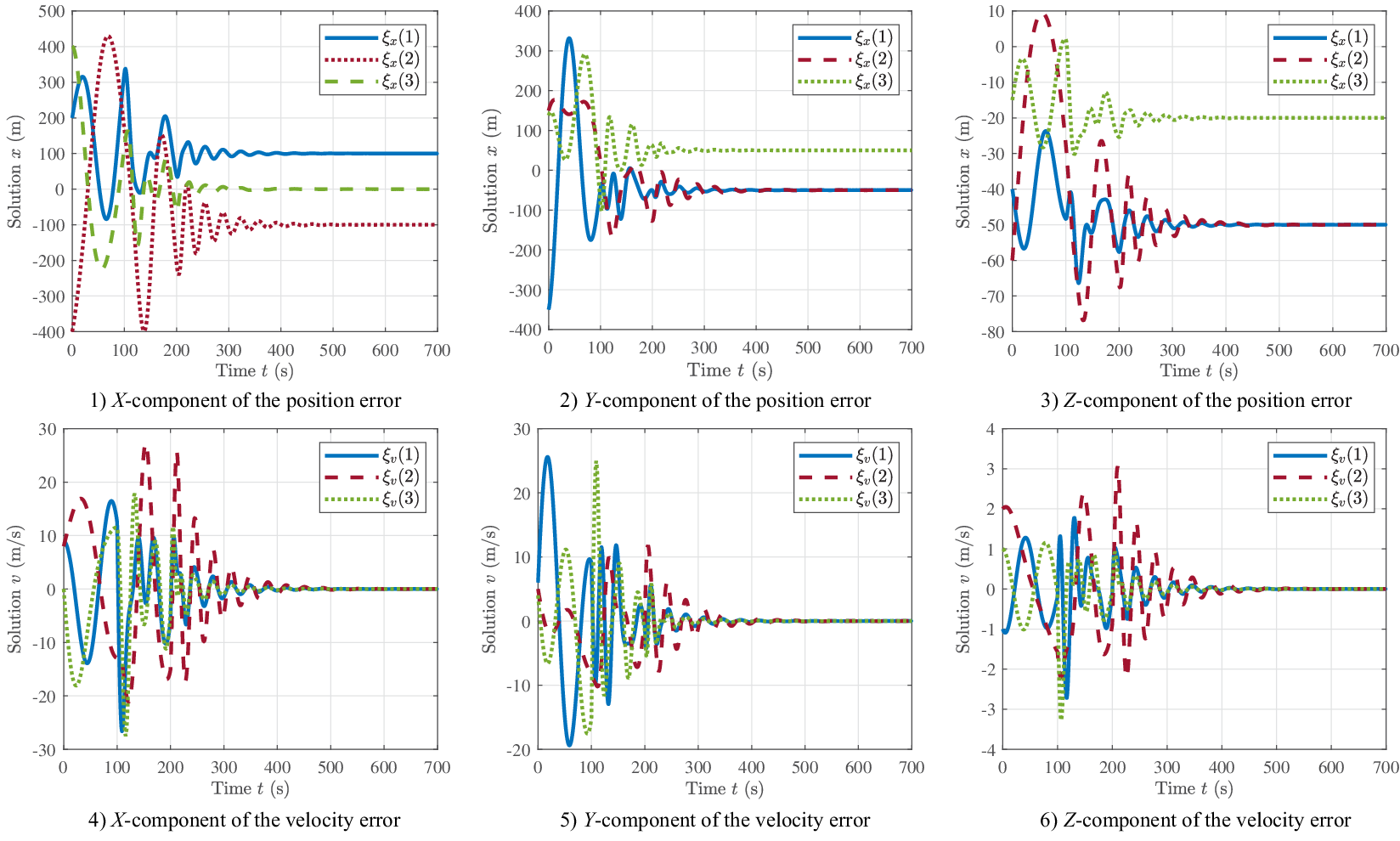}
    	}
    	\subfloat[The 3D trajectories]{
    		\label{Fig: case1-3d}
    		\raisebox{0.55\height}{\includegraphics[width=0.25\linewidth]{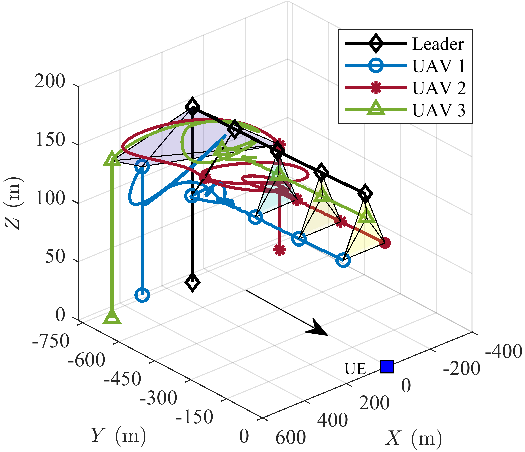}}
    	}
    	\caption{The error dynamics and 3D trajectories of a multi-agent system consisting of four UAVs in the case of static UE located at the origin.}
    	\label{Fig: case1}
    \end{figure*}
	
	According to Theorem~\ref{Thm: formation}, we set $c \geq 0.002$ (sepcifically, $c = 0.002$ with time $t \leq 100$, $c = 0.02$ with $100 < t \leq 200$, and $c = 0.1$ with $t \geq 200$) in the pertaining simulation experiments. It is observed from Fig.~\ref{Fig: case1-xv} that each component of the position error for each UAV converges to a constant (cf. the upper panel) while each component of the velocity error of each UAV (cf. the lower panel) converges to zero as the time $t$ goes from $0$ to about $500$ \si{s}, and this convergence time decreases with higher UAV speed. As a result, we infer that the trajectories of all UAVs in this system achieve second-order leader-follower flight formation control. For intuitive understanding, Fig.~\ref{Fig: case1-3d} shows the complete trajectories of four UAVs, where the leader first flies to the designated height $150$~\si{m}. Then, the followers randomly take off from the ground and receive the leader's control to form an ideal formation. Afterward, they form a steady swarm and fly to the target locations predetermined by the location of the static UE.

	\section{UAV Swarm Tracking for Mobile UEs} \label{sec: control2}
	For mobile UEs, the UAV swarm must trace them to ensure their effective QoS. This section develops an impulsive formation tracking strategy that enables the UAV swarm to pursue a mobile UE. In this case, the multi-agent system consists of a UAV swarm and a mobile UE, and its dynamics differ from the UAV swarm studied in the preceding section.
	
	\subsection{The Dynamics of Mobile UEs and UAV Swarm} \label{sec: control2-a}
	By using a similar approach as in the Subsection~\ref{sec: control1-a}, the dynamics of a mobile UE can be described as 
	\begin{subequations} \label{Eq: C2-gUE}
		\begin{align}
			\dot{\bm{x}}_{g} (t) & = {\bm{v}_{g}} (t), \label{Eq: C2-gUE-a} \\
			\dot{\bm{v}}_{g} (t) & = {\bm g} \left(t, \bm{x}_{g} (t), \bm{v}_{g} (t)\right), \label{Eq: C2-gUE-b}
		\end{align}
	\end{subequations}
	where $\bm{x}_{g} (t) \in \mathbb{R}^{m}$ in \eqref{Eq: C2-gUE-a} denotes the position of the target mobile UE and its first-order differential $\bm{v}_{g} (t)$ refers to the velocity, and the function ${\bm g} \left(\cdot\right)$ in \eqref{Eq: C2-gUE-b}  represents the intrinsic nonlinear dynamics of the mobile UE.
	
	Given a UAV swarm consisting of one leader and $n$ followers, we adopt the impulsive control strategy to reduce their communication cost. Accordingly, the impulsive moments satisfy 
	\begin{equation}
		0 \leq t_{0} < t_{1} < \cdots < t_{k} < \cdots, \quad \lim_{k \to \infty} t_{k} = + \infty,
	\end{equation}
	and the impulsive interval $\Delta t_{k} \triangleq t_{k} - t_{k - 1} \leq \tau < \infty$, for all $k \in \mathbb{N}_{+}$. For descriptive purposes, we next focus on the time interval $(t_{k-1}, t_{k}]$, and the resulting conclusion applies to the other time intervals. On the one hand, for any time instant $t \in (t_{k-1}, t_{k})$, i.e., $t \neq t_{k}$, by recalling the theory of multi-agent systems, the dynamics of the UAV swarm can be described as
	\begin{subequations}\label{Eq: C2-general}
		\begin{align}
			\dot{\bm{x}}_{0} (t) & = {\bm{v}_{0}} (t), \label{Eq: C2-leader-a} \\
			\dot{\bm{v}}_{0} (t) & = {\bm f}_{0} \left(t\right), \label{Eq: C2-leader-b} \\
			\dot{\bm{x}}_{i} (t) & = \bm{v}_{i} (t), \label{Eq: C2-leader-c} \\
			\dot{\bm{v}}_{i} (t) & = {\bm f}_{i} \left(t\right) + \sum_{j = 1}^{n} a_{i j}  \left[c \left(\bm{x}_{j} (t) - {\bm{x}}^{\ast}_{j} - \bm{x}_{i} (t)  + {\bm{x}}^{\ast}_{i}\right) \right. \nonumber \\
			& \quad  \left. + c \left(\bm{v}_{j} (t) - \bm{v}_{i} (t)\right)\right], \label{Eq: C2-leader-d}
		\end{align}
	\end{subequations}
	where the variables have similar meanings to those in \eqref{Eq: C1-general}, except that \eqref{Eq: C2-leader-d} differs significantly from \eqref{Eq: C1-follower-b} because of distinct control strategies. 
	
	On the other hand, at an impulsive moment, e.g., $t = t_{k}$, the jump of position and velocity of the agents at the impulsive moment $t_{k}$ can be computed and given by
	\begin{subequations}\label{Eq: C2-impulse}
		\begin{align}
			\Delta \bm{x}_{0} (t) & = - \ell_{x}^{0} (t) \left(\bm{x}_{0} (t) - \bm{x}_{g} (t)\right),
			\label{Eq: C2-impulse-a} \\
			\Delta \bm{v}_{0} (t) & = - \ell_{v}^{0} (t) \left(\bm{v}_{0} (t) - \bm{v}_{g} (t) \right),
			\label{Eq: C2-impulse-b} 
		\end{align}
		where $\ell_{x}^{0} (t)$ in \eqref{Eq: C2-impulse-a} and $\ell_{v}^{0} (t)$ in \eqref{Eq: C2-impulse-b} denote the control strength for position and velocity of the leader, respectively;  
		\begin{align}
			\Delta \bm{x}_{i} (t) & = - \ell_{x}^{i} (t) \left(\bm{x}_{i} (t) - \bm{x}^{\ast}_{i} - \bm{x}_{g} (t)\right), \label{Eq: C2-impulse-c} \\
			\Delta \bm{v}_{i} (t) & = - \ell_{v}^{i} (t) \left(\bm{v}_{i} (t) - \bm{v}_{g} (t)\right), \label{Eq: C2-impulse-d} \\
			\bm{0} & \leq {\Delta \bm{x}}_{i} (t) \leq \bm{\delta}_{\bm{x}}^{\max}, 	\label{Eq: C2-impulse-e} \\			
			\bm{0} & \leq {\Delta \bm{v}}_{i} (t) \leq \bm{\delta}_{\bm{v}}^{\max}, 	\label{Eq: C2-impulse-f}
		\end{align}
	\end{subequations}
	for all $i = 1, \cdots, n$, where $\ell_{x}^{i} (t)$ in \eqref{Eq: C2-impulse-c} and $\ell_{v}^{i} (t)$ in \eqref{Eq: C2-impulse-d} denote the control strength for position and velocity of the $i^{\text{th}}$ follower; $\bm{\delta}_{\bm{x}}^{\max}$ in \eqref{Eq: C2-impulse-e} and $\bm{\delta}_{\bm{v}}^{\max}$ in \eqref{Eq: C2-impulse-f} represent the maximum displacement and velocity at the impulsive moment $t$, respectively, and the inequality ``$\leq$'' in \eqref{Eq: C2-impulse-e}-\eqref{Eq: C2-impulse-f} holds in the element-wise sense. In addition, for the sake of mathematical continuity but without loss of generality, we assume  
	\begin{equation}
		\bm{v}_{0} (t_{0}^{+}) = \bm{v}_{0} (t_{0}) \text{ and } \bm{v}_{i} (t_{0}^{+}) = \bm{v}_{i} (t_{0}), \ \forall t_{0} \geq 0.
	\end{equation}
	
	\subsection{UAV Swarm Tracking} \label{sec: control2-b}	
	To ensure effective tracking, the UAV swarm must be controlled in a 3D space around the pursued mobile UE, i.e., in a state admissible set $\mathcal{Q}$. Moreover, we define another set corresponding to $\mathcal{Q}$ as $\mathcal{Q}_{0} = \left\{\bm{x} \in \mathcal{Q}_{0} \mid \mathcal{B} \left(\bm{x}, r\right) \in \mathcal{Q}\right\}$, where $\mathcal{B} \left(\bm{x}, r\right)$ is a open Eucliden ball with center~$\bm{x}$ and radius~$r$. In our applications, $r$ is determined by the radius of the circumcircle of the UAV formation topology. Clearly, $\mathcal{Q}_{0}$ is a proper subset of $\mathcal{Q}$. Based on the above definitions, we assign the values of $\ell_{x}^{i}(t)$ and $\ell_{v}^{i}(t)$ such that they have the same values in \eqref{Eq: C2-impulse-c}-\eqref{Eq: C2-impulse-d} at the impulsive moment $t_k$, computed by
	\begin{equation}\label{Eq: C2-l}
		\ell^{i}(t_{k}) =
		\begin{cases}
			\min\limits_{i} \left\{\dfrac{\bm{\delta}_{\bm{x}}^{\max}}{\bm{x}_{g} (t_{k}) - \bm{x}_{i}(t_{k}) + \bm{x}_{i}^{\ast}},  \dfrac{\bm{\delta}_{\bm{v}}^{\max}}{\bm{v}_{g}(t_{k})) - \bm{v}_{i}(t_{k})} \right\}, & \\   \qquad \qquad \qquad \qquad \text{if a UAV is outside of} \ \mathcal{Q}_{0}; \\
			0, \qquad \qquad \qquad \quad \text{otherwise}.
		\end{cases}
	\end{equation}
	It is noteworthy that \eqref{Eq: C2-l} is also applicable to the leader with $\bm{x}^{\ast}_{0} = \bm{0}$. 
	
	By virtue of \eqref{Eq: C2-general}-\eqref{Eq: C2-l}, the tracking errors can be explicitly computed as:  
	\begin{subequations}\label{Eq: C2-error_0}
		\begin{align}
			\bm{\zeta}_{\bm{x}}^{L} (t) & =  \bm{x}_{0} (t) - \bm{x}_{g} (t),
			\label{Eq: C2-error-leader-x}	\\
			\bm{\zeta}_{\bm{v}}^{L} (t) & = \bm{v}_{0} (t) - \bm{v}_{g} (t)
			\label{Eq: C2-error-leader-v}	\\
			\bm{\zeta}_{\bm{x}} (t) & = \left[\begin{smallmatrix} (\bm{x}_{1} (t) - \bm{x}_{g} (t) - \bm{x}_{1}^{\ast})^{\text T}, & \cdots, & (\bm{x}_{n} (t) - \bm{x}_{g} - \bm{x}_{n}^{\ast})^{\text T} \end{smallmatrix}\right]^{\text T},
			\label{Eq: C2-error-x} \\
			\bm{\zeta}_{\bm{v}} (t) & = \left[\begin{smallmatrix} (\bm{v}_{1} (t) - \bm{v}_{g} (t))^{\text T}, & \cdots, & (\bm{v}_{n} (t) - \bm{v}_{g})^{\text T} \end{smallmatrix}\right]^{\text T}.
			\label{Eq: C2-error-v} 
		\end{align}
		For ease of further proceeding, combining \eqref{Eq: C2-error-leader-x} (resp., \eqref{Eq: C2-error-x}) and \eqref{Eq: C2-error-leader-v} (resp., \eqref{Eq: C2-error-v}), we define 
		\begin{align} 
			\bm{\epsilon}_{g}^{L} (t) & \triangleq \begin{bmatrix} \left(\bm{\zeta}_{\bm{x}}^{L}\right)^{\text T} (t), & \left(\bm{\zeta}_{\bm{v}}^{L}\right)^{\text T} (t) \end{bmatrix}^{\text T}, 
			\label{Eq: C2-error-leader-xv}\\
			\bm{\epsilon}_{g}^{F} (t) & \triangleq \begin{bmatrix} \bm{\zeta}_{\bm{x}}^{\text T} (t), & \bm{\zeta}_{\bm{v}}^{\text T} (t) \end{bmatrix}^{\text T},
			\label{Eq: C2-error-xv}
		\end{align}	
	\end{subequations}
	where the superscripts ``{\it L}'' and ``{\it F}'' refer to ``Leader'' and ``Follower'', respectively.
	
	In light of \eqref{Eq: C2-gUE}, \eqref{Eq: C2-general}, \eqref{Eq: C2-impulse}, and \eqref{Eq: C2-error_0}, the error dynamics for any time instant $t \in (t_{k-1}, t_{k}]$ can be computed and after some algebraic manipulations, we obtain
	\begin{subequations} \label{Eq: C2-error}
		\begin{align}
			\dot{\bm{\epsilon}}_{g}^{L} (t) & = \bm{G}^{L} {\bm \epsilon}_{g}^{L} (t) +
			\left[\begin{smallmatrix}
				\bm{0}_{m} \\ \bm{f} \left(t\right) - \bm{g} \left(t\right)
			\end{smallmatrix}\right], \quad t \neq t_{k},
			\\
			\dot{\bm{\epsilon}}_{g}^{F} (t) & = \bm{G}^{F} {\bm \epsilon}_{g}^{F} (t) +
			\left[\begin{smallmatrix}
				\bm{0}_{mn} \\ {\bm F} \left(t\right) - \mathbbm{1}_{n} \otimes {\bm g} \left(t\right)
			\end{smallmatrix}\right], \quad t \neq t_{k},
			\\
			\Delta {\bm{\epsilon}}_{g}^{L} (t_{k}) & =  \left[\begin{smallmatrix}
				- \ell^{0}(t_{k})  	& 0  \\
				0						& - \ell^{0} (t_{k})
			\end{smallmatrix}\right] \otimes \bm{I}_{m} \bm{\epsilon}_{g}^{L} (t_{k}) \triangleq \bm{H}_{k}^{L} \bm{\epsilon}_{g}^{L} (t_{k}), \\
			\Delta {\bm{\epsilon}}_{g}^{F} (t_{k}) & =  \bm{H}_{k}^{F} \bm{\epsilon}_{g}^{F} (t_{k}),
		\end{align} 
	\end{subequations}
	where the matrices ${\bm G}^{L}$ and ${\bm G}^{F}$ are defined as
	\begin{equation}
		\bm{G}^{L} \triangleq \left[\begin{smallmatrix} 0 & 1 \\ 0 & 0 \end{smallmatrix}\right]  \otimes \bm{I} _{m}, {\text{ and }}
		\bm{G}^{F} \triangleq \left[\begin{smallmatrix} \bm{0}_{n} & \bm{I}_{n}  \\ c \, \bm{L}   & c \, \bm{L} \end{smallmatrix}\right]  \otimes \bm{I} _{m},
	\end{equation}
	with $\bm{L}$ being the graph Laplacian matrix of the followers' communication; $\bm{F} (t) \triangleq [\bm{f}_{1}^{\text T} (t),  \cdots,  \bm{f}_{n}^{\text T} (t)]^{\text T}$, and
	\begin{equation*}
		\bm{H}_{k}^{F} \triangleq
		\left[\begin{smallmatrix}
			{\text{Diag}}\left\{- \ell^{1}(t_{k}), \cdots, - \ell^{n}(t_{k})\right\}   	& \bm{O}_{n}  \\
			\bm{O}_{n}					& 	{\text{Diag}}\left\{- \ell^{1}(t_{k}), \cdots, - \ell^{n}(t_{k})\right\}
		\end{smallmatrix}\right]  \otimes \bm{I}_{m}.
	\end{equation*}

    \begin{figure}[t]
    	\centering
    	\includegraphics[width=0.95\linewidth]{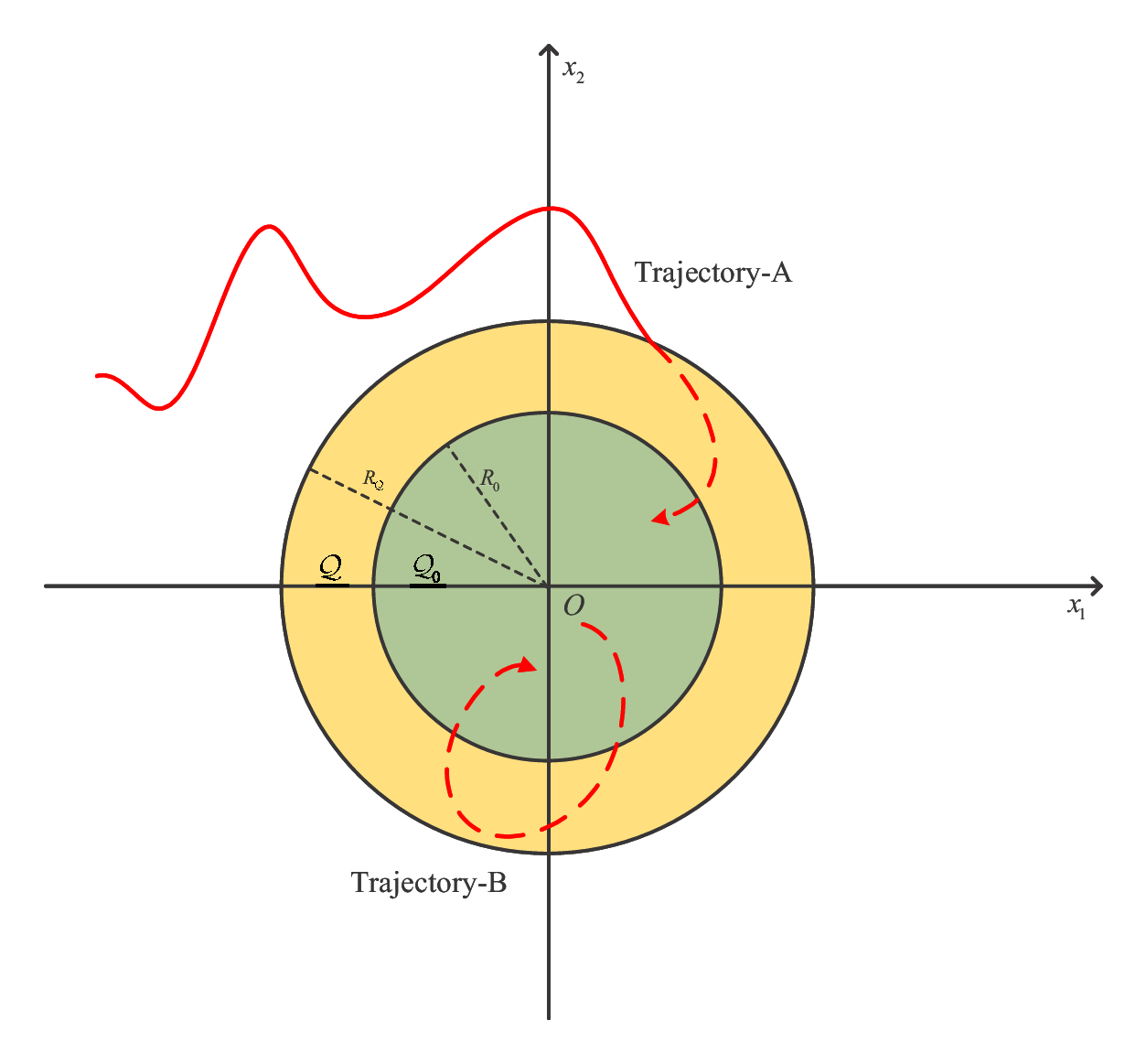}
    	\vspace{-10pt}
    	\caption{A sketch of solution trajectories of the error dynamics \eqref{Eq: C2-error}, where the $\mathcal{Q}_{0}$ (the green region) denotes the initial state set and $\mathcal{Q}$ (the golden region) is the corresponding state admissible set.}		
    	\label{Fig: stateset}
    \end{figure}
	
	We can now formalize the main result for UAV swarm tracking as follows.
	\begin{theorem} \label{Thm: tracking}
		Given Assumptions~\ref{Asump: graph} and \ref{Asump: Lipschitz}, if there exists $\rho > 1$ such that, for all $k$,
		\begin{equation}\label{Eq: C2-thm1}
			\eta (t_{k} - t_{k - 1}) + \ln(\rho \beta_{k}) < 0,
		\end{equation}
		where 	
		\begin{align}
			\eta & \triangleq \max \left\{\left[\left({\bm G}^{F}\right)^{\rm T} + {\bm G}^{F}\right], 1\right\} \nonumber \\
			& \quad + 2 \left(\max \left\{\left\|{\bm F}(t)\right\|_{\infty}, \left\|{\bm f}_{0} (t)\right\|_{\infty} \right\}\right) + 2 \left\|{\bm g} (t)\right\|_{\infty}, 
			\label{Eq: C2-thm1-parameter1} \\
			\beta & \triangleq \max \left\{\lambda_{\max} \left(\bm{I}_{m n}  + {\bm H}_{k}^{F}\right), \lambda_{\max} \left(\bm{I}_{2 m}  + {\bm H}_{k}^{L} \right)\right\},
			\label{Eq: C2-thm1-parameter2}
		\end{align}
		then, the multi-agent tracking system keeps the required formation and remains in the state admissible set $\mathcal{Q}$ corresponding to $\mathcal{Q}_{0}$ when $t_{k}$ is sufficiently large. Moreover, the maximum radius of $\mathcal{Q}$ is determined by
		\begin{align} \label{Eq: C2-thm2}
			R_{\mathcal{Q}} & = R_{0} + \max \bigg\{ \left(\|\bm{f}_{0} (t)\|_{\infty} + \|\bm{g}(t)\|_{\infty}\right) \tau,  \nonumber \\
			& \quad R_{0} \tau +\frac{1}{2} \left(\|\bm{f}_{0} (t)\|_{\infty} + \|\bm{g}(t)\|_{\infty}\right) \tau^{2} \bigg\} + \varepsilon,
		\end{align}
		where $R_{0}$ is the radius of $\mathcal{Q}_{0}$, which means the maximum of the UAVs' velocities and the distances between each UAV and the mobile UE; $\tau \triangleq \max_{k} \Delta t_{k}$, and $\varepsilon$ is infinitesimal.
	\end{theorem}
	
	\begin{IEEEproof}
		For brevity, we sketch the main idea of the proof as follows. As illustrated in Fig.~\ref{Fig: stateset}, we want to keep the tracking error in an acceptable neighborhood (e.g., the set $\mathcal{Q}_{0}$ in Fig.~\ref{Fig: stateset}) by using the impulsive control strategy. However, the dynamics of trackers (UAVs) jump at the impulsive moment $t_{k}$ due to the characteristics of the control strategy. To ensure the stability of control variables in \eqref{Eq: C2-impulse}, it is necessary to prove that the initial error value outside $\mathcal{Q}_{0}$ can reach $\mathcal{Q}_{0}$ through the control (e.g., the ``Trajectory-A'' in Fig.~\ref{Fig: stateset}), and the maximum error $\epsilon(t_{2})$ of any agent with $\epsilon(t_{1}) \leq R_{0}$ never exceeds $\mathcal{Q}$ (e.g., the ``Trajectory-B'' in Fig.~\ref{Fig: stateset}). Consequently, the maximum radius of $R_{\mathcal{Q}}$ can be expressed as the minimum bound of displacement and velocity at each $\Delta t_{k}$. For more details of full proof, please refer to Appendix \ref{Appendix: C2}.
	\end{IEEEproof}
	
	Finally, it remains to state that the tracking control strategy developed above can also achieve formation translation, rotation, scale, and other geometric deformations by combining with further control such as the affine formation control \cite{Zhao2018}. 
	
	\subsection{Case Study} \label{sec: control2-c}	
	
	To illustrate the effectiveness of the swarm tracking control strategy, like the preceding Subsection~\ref{sec: control1-c}, we consider a nonlinear system consisting of four UAVs ($i = 0, 1, 2, 3$) and one mobile UE in the pertaining simulation experiments. In this case, it is more important to consider the tracking error between the UAV swarm and the UE than the control error within the swarm. Assume the initial position of the leader UAV is in a circular region $\mathcal{Q}$, whose center is at the initial position of the target mobile UE and the radius is $10$~\si{\m}; and $\mathcal{Q}_{0}$ is the set just including the position of the target mobile UE as an extreme case. Furthermore, assume the mobile UE moves along a zigzag path at a constant speed of $10$ \si{m/s}.
	
    \begin{figure*}
    	\captionsetup[subfigure]{margin=5pt}
    	\subfloat[The position and velocity errors]{
    		\label{Fig: case2-xv}
    		\includegraphics[width=0.7\linewidth]{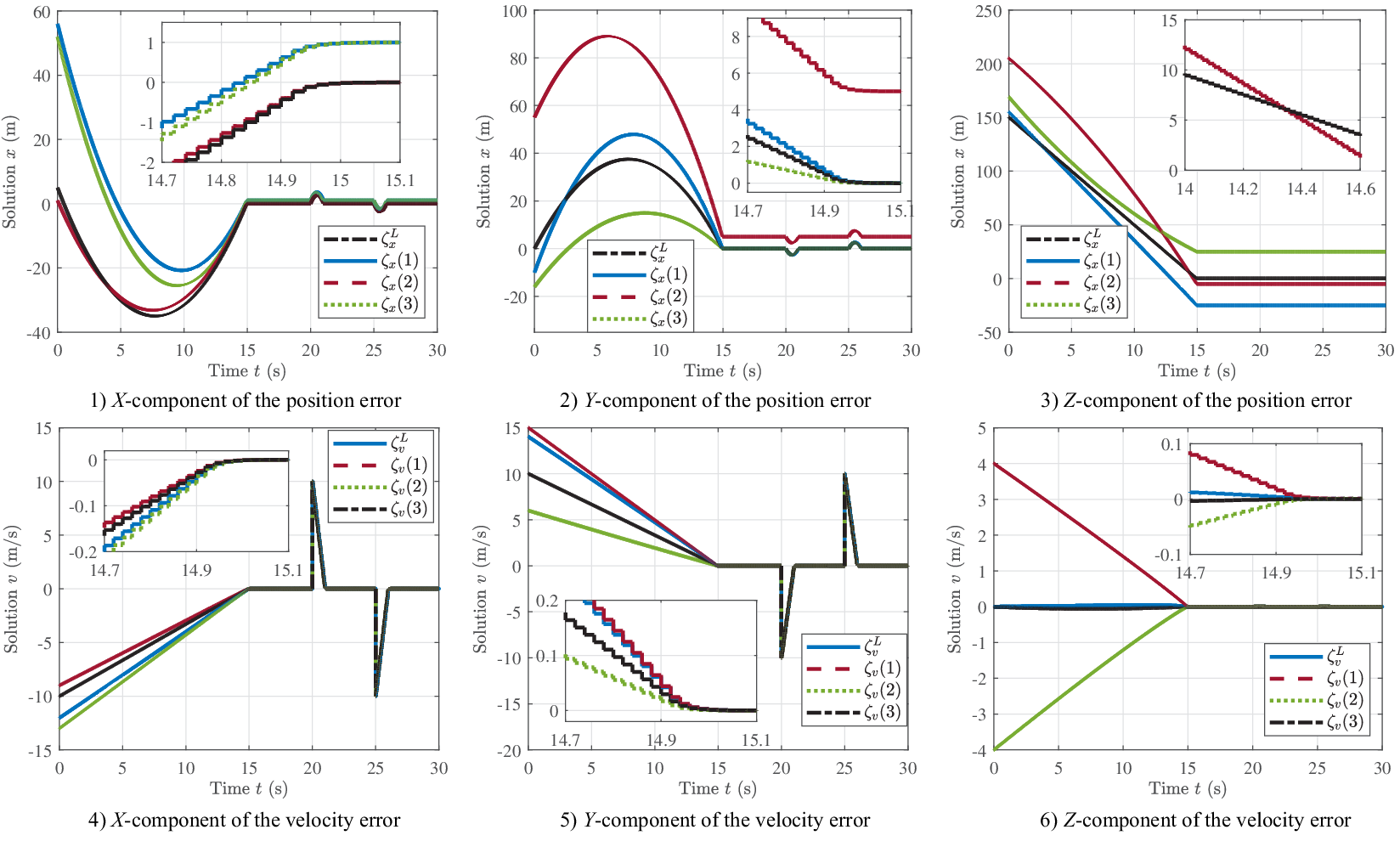}
    	}
    	\subfloat[The 3D trajectories]{
    		\label{Fig: case2-3d}
    		\raisebox{0.55\height}{\includegraphics[width=0.25\linewidth]{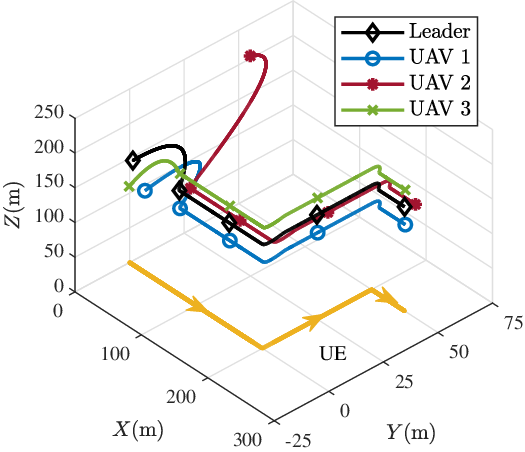}}
    	}
    	\caption{The error dynamics and 3D trajectories of a multi-agent system consisting of four UAVs in the case of mobile UE.}
    	\label{Fig: case2}
    \end{figure*}
	
	Given the equidistant impulsive interval $\Delta t_{k} = \tau = 0.02$~\si{\s}, we obtain $\eta = 23$ and $\beta = 0.5$ by use of \eqref{Eq: C2-thm1-parameter1} and \eqref{Eq: C2-thm1-parameter2}, respectively. Thus there exists $\rho \geq 1.3$, which satisfies the inequality~\eqref{Eq: C2-thm1} in Theorem~\ref{Thm: tracking}. For illustration purposes, Fig.~\ref{Fig: case2-xv} demonstrates that each component of the position error between the UAVs and the mobile UE converges to a constant (cf. the upper panel) while each component of the velocity error between the UAVs and the mobile UE (cf. the lower panel) converges to zero as the time $t$ goes from $0$ to $30$ \si{s}. These observations show that the tracking system keeps the required formation and effective tracking of the mobile UE under our impulsive control strategy. For intuitive understanding, Fig.~\ref{Fig: case2-3d} shows a complete trajectory of the tracking system, which illustrates the effective tracking capability of the UAV swarm. It is noteworthy that the error curves in Fig.~\ref{Fig: case2-xv} show some jitters at $20$~\si{s} and $25$~\si{s} due to the changed movement direction of UE at these two instants, respectively. Even so, these error curves become stabilized within $2$~\si{s}.

	\section{Analytical Performance Analysis} \label{sec: performance}
	After making the formation control and tracking strategies of a UAV swarm in previous sections, a CoMP set of collaborative UAVs can consistently serve static or mobile terrestrial UEs. So, in this section, we analyze the coverage probability and ergodic rate of a typical UE to evaluate the transmission performance of the UAV CoMP.
	
	\subsection{Signal Model}	
	As mentioned in Section~\ref{sec: control2-c}, the 3D space is considered an inertial coordinate system, which follows the right-hand rule. Naturally, the ground can be regarded as the $X$-$Y$ plane of this 3D system. Without loss of generality, we analyze the performance of a typical UE located at the origin $\bm{o} = (0, 0, 0)$ of the space according to the Slivnyak–Mecke theorem \cite[Thm. 1.4.5]{Baccelli2009}. The 2D and 3D Euclidean distances from the $i^{\rm th}$ serving UAV to the origin are denoted by $r_{i}$ and $d_{i} = \sqrt{r_{i}^{2} + h_{i}^{2}}$, respectively. Assume that each UAV in the network is equipped with multi-antennas while each UE has a single antenna. Consequently, the received signal at a typical UE can be computed as
	\begin{align}
		y_{0} & = \sum\limits_{i \in \Phi_{0}} ({P_{i}} {d_{i}^{- \alpha}})^{\frac{1}{2}} {\bm{h}_{i}^{\text H}} {\bm{\omega_{i}}} {\bm{s}_{0}}  \nonumber\\
		& \quad + \sum\limits_{j \in \Phi_{B} \backslash \Phi_{0}} ({P_{j}} {d_{j,0}^{- \alpha}})^{\frac{1}{2}} {\bm{h}_{j,0}^{\text H}}{\bm{\omega_{j}}} {\bm{s}_{j}} + n,
		\label{Eq: channel}
	\end{align}
	where $P_{i} \in \mathbb{R}$ denotes the transmit power of the $i^{\rm th}$ UAV; $d_{i}^{- \alpha}$ indicates the large-scale path loss from the $i^{\rm th}$ UAV to a typical UE, with the path-loss exponent $\alpha \ge 2$; $\bm{h}_{i}$ refers to the small-scale fading; $\bm{\omega_{i}} \triangleq \bm{h_{i}}/\|\bm{h_{i}}\|$ is a maximum ratio transmission (MRT) vector at the $i^{\rm th}$ UAV, and $\bm{s}_{0}$ is the transmit signal of UAVs in the CoMP set $\Phi_{0}$. Also, in the second term on the right-hand-side of \eqref{Eq: channel}, the parameter $P_{j}$ means the transmit power of the $j^{\rm th}$ UAV; $d_{j, 0}$ is the Euclidean distance from the $j^{\rm th}$ UAV as an interferer to a typical UE; $\bm{h}_{j, 0}$ stands for the small-scale fading between the $j^{\rm th}$ UAV and a typical UE, and $\bm{s}_{j}$ refers to the interfering signal transmitted by the $j^{\rm th}$ UAV. Finally, the last term $n$ of \eqref{Eq: channel} stands for a circularly-symmetric additive white Gaussian noise at a typical UE, with mean zero and variance $\sigma^{2}$. It remains to note that intra-cell interference is not accounted for in \eqref{Eq: channel} because it can be effectively mitigated in practice by using techniques such as successive interference cancellation (SIC) and orthogonal frequency division multiple access (OFDMA) for static UEs. In the case of mobile UEs, the cooperative UAVs can form a distributed multiple-input multiple-output (MIMO) antenna system and generate multiple dynamic beams, each serving one mobile UE. This paper focuses on single-user rather than multi-user cases for simplicity without loss of generality.\footnote{For the cases of multiple static UEs or multiple mobile UEs but with similar movement patterns, like a vehicle platoon, their locations can be combined into a single location information such that the control strategy of UAVs for multiple UEs can be designed similarly to the single UE case. However, if various UEs move in diverse directions, multiple UAV swarms must be used to track them each.}
	
	Since the UAVs under consideration have the same configuration, their transmit powers are assumed to be identical and normalized to unity. Also, as the UE performance is generally interference-limited, the noise term in \eqref{Eq: channel} is negligible. Consequently, the signal-to-interference ratio (SIR) at a typical UE can be computed as 
	\begin{equation}
		{\text{SIR}} \triangleq \frac{S}{I}
		= \frac{\left| \sum\limits_{i \in \Phi_{0}} ({d_{i}^{- \alpha}})^{\frac{1}{2}} \|\bm{h}_{i}\| \right|^{2}} {\sum\limits_{j \in \Phi_{B} \backslash \Phi_{0}} \left| ({d_{j,0}^{-\alpha}})^{\frac{1}{2}} {\bm{h}_{j, 0}^{\text H}} \frac{\bm{h_{j}}}{\|\bm{h_{j}}\|} \right|^{2}}.
		\label{Eq: SIR}
	\end{equation}
	As air-to-ground communication links may experience various fading scenarios, we exploit the Nakagami-$m$ fading model to capture a large class of fading environments. Accordingly, the power gain of small-scale fading between a typical UE and a UAV follows the Gamma distribution with PDF given by
	\begin{equation} 
		f_{\|\bm h\|}(x) = \frac{2 m^{m} x^{2 \si{m} - 1}} {\Gamma(m) \Omega^{m}} \exp \left(-\frac{m}{\Omega} x^{2}\right), \ m \geq 0.5.
		\label{Eq: nakagami}
	\end{equation}	
	
	To evaluate the efficiency of CoMP transmission, we investigate two performance metrics, i.e., the coverage probability and ergodic rate. Mathematically, given an outage threshold on the received SIR at a typical UE, say $\gamma_{\rm th}$, the coverage probability is defined as
	\begin{equation}
		\mathcal{P} \left(\gamma_{\rm th}\right) = 1 - {\text{Pr}}\{\text{SIR} \leq \gamma_{\rm th}\},
		\label{Eq: def-cp}
	\end{equation}
	and by the seminal Shannon formula, the ergodic rate of a typical UE is calculated by
	\begin{equation}
		\mathcal{R} = \mathbb{E}\left[\ln\left(1+\text{SIR}\right)\right].
		\label{Eq: def-adr}
	\end{equation}
	
	\subsection{The Statistics of Received Signal and Interference}
	Since $\Phi_{x}$ is a 2D PPP, the horizontal distance between a typical UE at the origin and its $i^{\rm th}$ nearest UAV projection onto the plane is distributed as \cite[Thm. 1]{Haenggi2005}:
	\begin{equation}
		f_{r_{i}}(r) = \frac{2(\lambda_{x} \pi)^{n}}{\Gamma(n)} r^{2 n - 1} \exp \left(-\lambda_{x} \pi r^{2}\right).
		\label{Eq: rn-pdf}
	\end{equation}
	Denote the PDF of UAV height $h$ as $f_{H}(h)$. Then, the PDF of $d_{i} = \sqrt{r_{i}^{2} + h^{2}}$ in \eqref{Eq: SIR} can be readily computed as
	\begin{align}
		f_{d_{i}}(x)
		& = \frac{(\lambda_{x} \pi)^{n} x}{\Gamma(n)} \int_{H_{\min}}^{H_{\max}} (x^{2} - h)^{n - 1} h^{- \frac{1}{2}} f_{H} \nonumber \\
		& \quad \times \left(h^{\frac{1}{2}}\right) \exp\left(\lambda_{x} (x^{2} - h)^{\frac{1}{2}}\right) \, {\rm d} h.
		\label{Eq: dn-pdf}
	\end{align}
	
	Recalling that there are four collaborative UAVs in the CoMP set $\Phi_{0}$ due to the modeling of Delaunay tetrahedralization as shown in Fig.~\ref{Fig: model}, in light of \eqref{Eq: SIR}, let the aggregated received signal strength $S \triangleq \left| T\right|^{2}$ with $T = \sum_{i = 1}^{4} d_{i}^{-\alpha/2}\|\bm{h}_{i}\|$. Although $\|\bm{h}_{i}\|$ and $d_{i}$ are independent, the exact PDF of $T$ is hard to obtain in closed-form. To address this issue, we employ the causal form of the central limit theorem \cite[p.235]{Papoulis1962} to approximate the sum of positive independent and not necessarily identically distributed random variables by a Gamma distribution. In particular, we have the following lemma to approximate the PDF of $T$.		
	
	\begin{lemma} \label{Lem: T-pdf}
		The PDF of $T \triangleq \sum_{i = 1}^{4} d_{i}^{-\alpha/2}\|\bm{h}_{i}\|$ can be approximately given by
		\begin{equation}
			f_{T}(x) \approx \frac{x^{\nu - 1}}{\theta^{\nu} \Gamma(\nu)} \exp \left(- \frac{x}{\theta}\right),
			\label{Eq: T-apdf}
		\end{equation}
		 where the parameters $\nu$ and $\theta$ are computed by \eqref{Eq: T-para1} and \eqref{Eq: T-para2}, respectively, shown on the bottom of the next page, 
		\begin{figure*}[!b]
			\normalsize
			\setcounter{mytempeqncnt}{\value{equation}}
			\setcounter{equation}{32}
			\vspace*{4pt}
			\hrulefill
			\begin{align}
				\nu & = \frac{\left(\sum\limits_{i = 1}^{4} \mathbb{E}\left[d_{i}^{- \frac{\alpha}{2}} \right]\right)^{2}} {m \left[\frac{\Gamma(m)}{\Gamma(m + \frac{1}{2})}\right]^{2} \left(\sum\limits_{i = 1}^{4} \mathbb{E}\left[d_{i}^{- \alpha} \right] + \sum\limits_{i = 1, i \neq j}^{4} \mathbb{E}\left[d_{i}^{- \frac{\alpha}{2}} \right] \mathbb{E}\left[d_{j}^{- \frac{\alpha}{2}} \right] \right) - \left(\sum\limits_{i = 1}^{4} \mathbb{E}\left[d_{i}^{- \frac{\alpha}{2}} \right]\right)^{2}},
				\label{Eq: T-para1}  \\
				\theta & = \frac{(m \Omega)^{\frac{1}{2}} \Gamma(m)} {\Gamma(m + \frac{1}{2}) \left(\sum\limits_{i = 1}^{4} \mathbb{E}\left[d_{i}^{- \frac{\alpha}{2}}\right]\right)} \left(\sum\limits_{i = 1}^{4} \mathbb{E}\left[d_{i}^{- \alpha} \right] + \sum\limits_{i = 1, i \neq j}^{4} \mathbb{E}\left[d_{i}^{- \frac{\alpha}{2}} \right] \mathbb{E}\left[d_{j}^{- \frac{\alpha}{2}} \right] \right) 
				- \frac{\Gamma(m + \frac{1}{2})}{\Gamma(m)} \left(\frac{\Omega}{m}\right)^{\frac{1}{2}}\left(\sum_{i = 1}^{4} \mathbb{E}\left[d_{i}^{- \frac{\alpha}{2}} \right]\right).
				\label{Eq: T-para2}
			\end{align}
			\setcounter{equation}{\value{mytempeqncnt}}
		\end{figure*}
		and the expectation of $d_{i}^{-\alpha}$ can be calculated by
		\begin{align} \label{Eq: T-para3}
			\setcounter{equation}{34}
			\mathbb{E}\left[d_{i}^{- \alpha}\right]
			& = \frac{1}{2}{(\lambda_{x} \pi)^{\frac{\alpha}{2}}} \int_{0}^{\infty} x^{- \frac{1}{2}} f_{H}\left(x^{\frac{1}{2}}\right) \nonumber\\
			& \quad \times \exp(\lambda_{x} \pi x) P_{\frac{\alpha}{2}}^{(i - 1)} (\lambda_{x} \pi x) \, {\rm d} x,
		\end{align}
		in which the function $P_{a}^{(n)}(x)$ denotes an $n^{\rm th}$-order polynomial of $x$ with parameter $a$, as explicitly defined in the notation section.
	\end{lemma}

	\begin{IEEEproof}
		See Appendix~\ref{Appendix: A1}. 
	\end{IEEEproof}
	
	Based on Lemma~\ref{Lem: T-pdf}, the CCDF and PDF of the strength of the desired signal, i.e., $S = \left| T\right|^{2}$, can be readily derived and expressed as
	\begin{equation*}
		F_{S}^{c}(x) \approx \frac{\Gamma\left(\nu, \frac{\sqrt{x}}{\theta} \right)}{\Gamma(\nu)}, 
	\end{equation*}
    and
    \begin{equation} \label{Eq: S-ccdf}
    	f_{S}(x) \approx \frac{x^{\frac{\nu - 2} {2}}}{2 \theta^{\nu} \Gamma(\nu)} \exp \left(- \frac{x^{\frac{1}{2}}}{\theta}\right),
    \end{equation}
	respectively.
	
	Now, we derive the PDF of the interference $I$ in \eqref{Eq: SIR}. Conditioned on the serving distance $r$ and using a similar derivation as in Appendix~\ref{Appendix: A1}, we have the following lemma.	
	
	\begin{lemma}\label{Lem: I-pdf}
		The PDF of $I \triangleq \sum_{j \in \Phi_{B} \backslash \Phi_{0}} {d_{j,0}^{-\alpha}} |g_{j}|^{2}$ can be accurately approximated by the Gamma distribution:
		\begin{equation}
			f_{I}(x) \approx \frac{x^{\nu^{\prime} - 1}}{{\theta^{\prime}}^{\nu^{\prime}} \Gamma(\nu^{\prime})} \exp \left(- \frac{x}{\theta^{\prime}}\right),
			\label{Eq: I-apdf}
		\end{equation}
		where
		\begin{align}
			\nu^{\prime}
			& = \frac{2 m^2 \pi \lambda_{x}  \left(\int_{H_{\min}}^{H_{\max}} \int_{0}^{\infty} x \left(x^{2} + h^{2}\right)^{- \frac{\alpha}{2}} f_{H}(h) \, {\rm d} x {\rm d} h \right)^{2}} {(m + 1) \Omega \int_{H_{\min}}^{H_{\max}} \int_{0}^{\infty} x \left(x^{2} + h^{2}\right)^{- \alpha} f_{H}(h) \, {\rm d} x {\rm d} h},
			\label{Eq: I-para1} \\
			\theta^{\prime}
			& = \frac{(m + 1) \Omega \int_{H_{\min}}^{H_{\max}} \int_{0}^{\infty} x \left(x^{2} + h^{2}\right)^{- \alpha} f_{H}(h) \, {\rm d} x {\rm d} h} {m \int_{H_{\min}}^{H_{\max}} \int_{0}^{\infty} x \left(x^{2} + h^{2}\right)^{- \frac{\alpha}{2}} f_{H}(h) \, {\rm d} x {\rm d} h}.
			\label{Eq: I-para2}
		\end{align}
	\end{lemma}
	
	\subsection{Performance Metrics}
	Now, we are in a position to summarize our main results regarding the coverage probability and ergodic rate of a typical UE in the following theorem.
	
	\begin{theorem}\label{Thm: typical}
		Given a prescribed outage threshold $\gamma_{\rm th}$, the coverage probability of a typical UE can be approximately computed by~\eqref{Eq: CP}, and the ergodic rate can be approximated by~\eqref{Eq: ADR}. Moreover, if $\nu + 2 \nu^{\prime} > 0$ and $\theta^{\prime} > 0$, then~\eqref{Eq: ADR} reduces to~\eqref{Eq: ADR1}. Equations~\eqref{Eq: CP}, \eqref{Eq: ADR}, and~\eqref{Eq: ADR1} are shown on the bottom of this page.
		\begin{figure*}[!b]
			\normalsize
			\setcounter{mytempeqncnt}{\value{equation}}
			\setcounter{equation}{39}
			\vspace*{4pt}
			\hrulefill
			\begin{align} 
				\mathcal{P} \left(\gamma_{\rm th}, \lambda_{x}, \alpha \right)
				& \approx \frac{1} {{\theta^{\prime}}^{\nu^{\prime}} \Gamma(\nu) \Gamma(\nu^{\prime})}\int_{0}^{\infty} x^{\nu^{\prime} - 1} \exp \left(- \frac{x} {\theta^{\prime}}\right) \Gamma\left(\nu, \frac{\left(\gamma_{\rm th} x\right)^{\frac{1}{2}}}{\theta} \right) {\rm d} x, 
				\label{Eq: CP}\\
				\mathcal{R} \left(\lambda_{x}, \alpha\right)
				& \approx \frac{1} {2 \theta^{\nu} {\theta^{\prime}}^{\nu^{\prime}} \Gamma(\nu) \Gamma(\nu^{\prime})} \left\{ \int_{0}^{\infty} z^{\frac{\nu - 2}{2}} \ln (1 + z) \right. \left. \times \left[\int_{0}^{\infty} x^{\frac{\nu + 2 \nu^{\prime} - 2}{2}} \exp \left(- \frac{(x z)^{\frac{1}{2}}}{\theta} - \frac{x} {\theta^{\prime}}\right) {\rm d} x \right]  {\rm d} z\right\},
				\label{Eq: ADR} \\
				\mathcal{R} \left(\lambda_{x}, \alpha\right)
				& \approx \frac{\left(\frac{2}{\theta^{\prime}}\right)^{- \frac{\nu + 2 \nu^{\prime}} {2}}} {2 \theta^{\nu} {\theta^{\prime}}^{\nu^{\prime}} \Gamma(\nu) \Gamma(\nu^{\prime})} \left[\int_{0}^{\infty}  z^{\frac{\nu - 2}{2}} \ln (1 + z) \exp \left(\frac{\theta^{\prime} z}{8 \theta^{2}}\right) D_{- (\nu + 2 \nu^{\prime})}\left(\sqrt{\frac{\theta^{\prime} z}{2 \theta^{2}}} \right) {\rm d} z\right]. \label{Eq: ADR1}
			\end{align}
			\setcounter{equation}{\value{mytempeqncnt}}
		\end{figure*}
	\end{theorem}
	
	\begin{IEEEproof}
		The coverage probability of a typical UE can be directly calculated as
		\begin{equation} \label{Eq: thm-cp}
			\setcounter{equation}{43}
			\mathcal{P} \left(\gamma_{\rm th}, \lambda_{x}, \alpha \right)
			= \mathbb{E}_{d} \left[{\rm Pr} \left(\text{SIR} > \gamma_{\rm th}\right) \right] 
			= \mathbb{E}_{d} \left[\int_{\gamma_{\rm th}}^{\infty} f_{\text{SIR}}(z) \, {\rm d} z \right].			
		\end{equation}
		In light of \eqref{Eq: S-ccdf} and \eqref{Eq: I-apdf}, we can derive the conditional PDF of the $\text{SIR}$ shown in \eqref{Eq: SIR}, given by
		\begin{align}
			f_{\text{SIR}}(z)
			& \approx \frac{z^{\frac{\nu - 2}{2}}} {2 \theta^{\nu} {\theta^{\prime}}^{\nu^{\prime}} \Gamma(\nu) \Gamma(\nu^{\prime})} \nonumber\\
			& \quad \times \int_{0}^{\infty} x^{\frac{\nu + 2 \nu^{\prime} - 2}{2}} \exp \left(- \frac{(x z)^{\frac{1}{2}}}{\theta} - \frac{x} {\theta^{\prime}}\right) {\rm d} x.
			\label{Eq: SIR1-apdf}
		\end{align}
		Substituting \eqref{Eq: SIR1-apdf} into \eqref{Eq: thm-cp} yields the intended \eqref{Eq: CP}. On the other hand, by definition, the ergodic rate can be computed as
		\begin{align}
			\mathcal{R} \left(\lambda_{x}, \alpha\right) 
				 = \mathbb{E} \left[\ln \left(1 + \text{SIR} \right)\right] 
				 = \int_{0}^{\infty} \ln \left(1 + \gamma \right) f_{\text{SIR}}(\gamma) \, {\rm d} \gamma. \label{Eq: thm-ADR}
		\end{align}
		Inserting \eqref{Eq: SIR1-apdf} into \eqref{Eq: thm-ADR} gives the desired \eqref{Eq: ADR}. Finally, \eqref{Eq: ADR1} can be obtained directly from \cite[Eq.(3.462.1)]{Gradshteyn2014} in case $\nu + 2 \nu^{\prime} > 0$ and $\theta^{\prime} > 0$.
	\end{IEEEproof}

	\section{Simulation Results and Discussions} \label{sec: simulation}
	In this section, we present and discuss Monte-Carlo simulation results, compared with the numerical results of the performance metrics computed as per the obtained analytical expressions in Section~\ref{sec: performance}. An extensive 3D wireless network with a circular coverage area with the radius $R = 3$ \si{\km} is assumed in the simulation experiments. Since we focus on LoS signal propagation conditions, the path loss exponent is set to $\alpha > 2$, and the fading parameter is set to $m = 2$ in the fast fading model \eqref{Eq: nakagami}.
	
	\subsection{The Accuracy of the Gamma Approximation in Lemmas~\ref{Lem: T-pdf} and \ref{Lem: I-pdf}}
	
\begin{figure}[!t]
		\centering
		\includegraphics[width=0.9\linewidth]{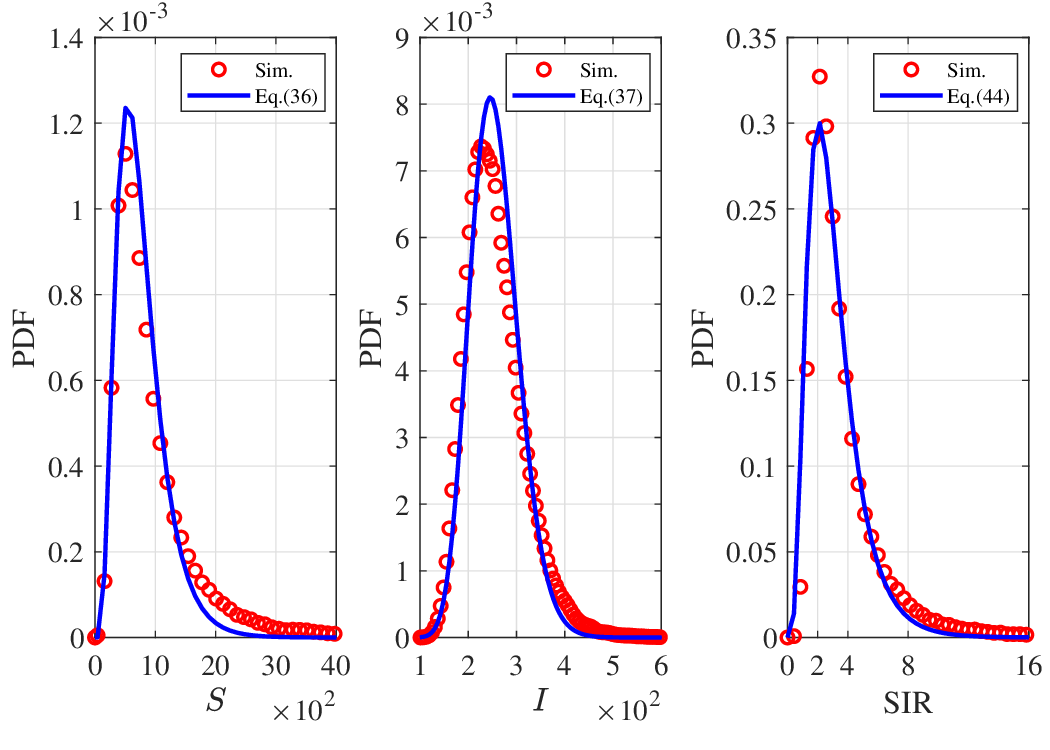} \vspace{-5pt}
		\caption{The PDF of the desired signal (left), the interference signal (middle), and their corresponding SIR (right) with $\alpha = 2.8$ and $\lambda_{x} = 16$ UAVs/\si{\km}$^2$.}
		\label{Fig: Gamma-approximation}
	\end{figure}

	Figure~\ref{Fig: Gamma-approximation} compares the numerical results computed as per our theoretical results with the simulation ones obtained in Monte-Carlo experiments, with the path-loss exponent $\alpha = 2.8$ and the intensity of UAV $\lambda_{x} = 16$ UAVs/\si{\km}$^2$. As shown in the left and middle panels of Fig.~\ref{Fig: Gamma-approximation}, the PDFs of the desired signal and unwanted interference computed by \eqref{Eq: S-ccdf} and \eqref{Eq: I-apdf} respectively almost agree with the simulation results. The slight deviation is mainly caused by two approximations: {\it a)} the Gamma approximation for the aggregated signal strength and, {\it b)} distance approximation, i.e., the selection of the serving UAV depends on the 2D distance between its projection and a typical UE rather than the actual 3D distance between them, as described in the penultimate paragraph in Section~\ref{sec: model}. On the other hand, as the PDF of SIR is applied to prove Theorem~\ref{Thm: typical}, the right panel of Fig.~\ref{Fig: Gamma-approximation} compares the numerical results computed by \eqref{Eq: SIR1-apdf} and the corresponding simulation ones, and they coincide very well with each other. 
	
	\subsection{Coverage Probability and Achievable Data Rate}
\begin{figure}[!t]
		\centering
		\includegraphics[width=0.9\linewidth]{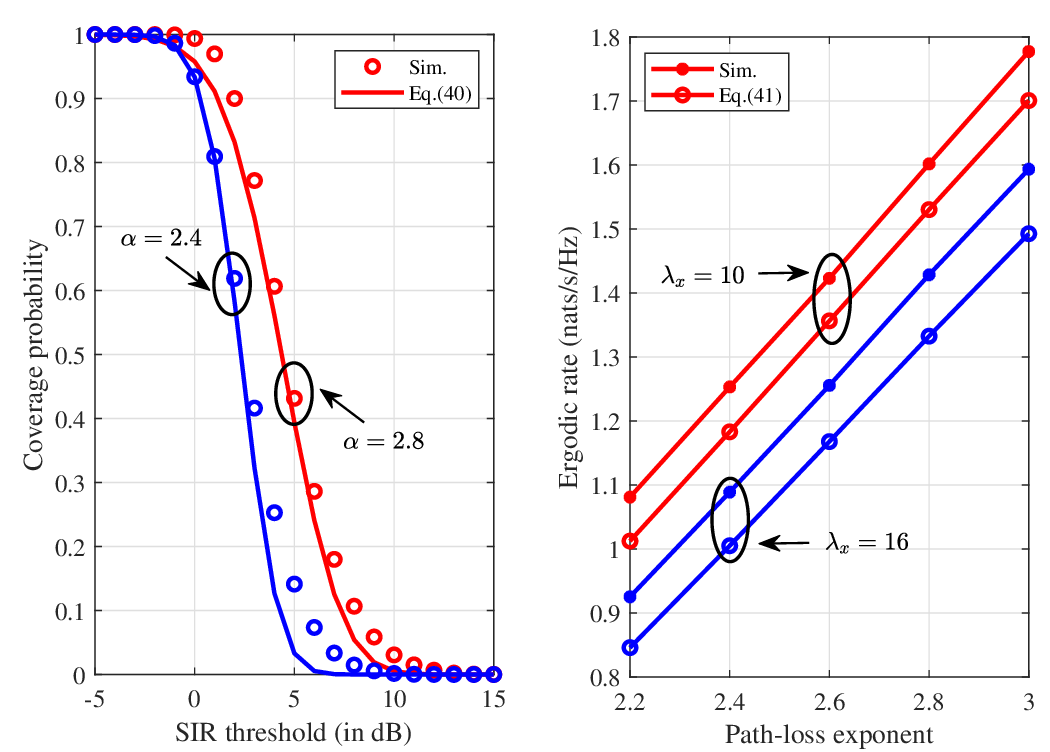} \vspace{-5pt}
		\caption{The coverage probability versus the SIR threshold (left), and the ergodic rate versus the path loss exponent (right) of a typical UE.}
		\label{Fig: performance1}
	\end{figure}

	To demonstrate the effectiveness of Theorem~\ref{Thm: typical}, Fig.~\ref{Fig: performance1} illustrates the coverage probability and ergodic rate for a typical UE. Its left panel depicts the coverage probability versus the SIR threshold, with the path-loss exponents $\alpha = 2.4, \, 2.8$. It is seen that the coverage probability increases with $\alpha$ for a fixed SIR threshold, as a larger $\alpha$ implies lower interference \cite{6291707}. Also, the numerical results computed as per \eqref{Eq: CP} in Theorem~\ref{Thm: typical} have similar trends with the simulation ones, which demonstrates the effectiveness of the approximate analysis.
	
	The right panel of Fig.~\ref{Fig: performance1} shows the ergodic rate versus the path loss exponent $\alpha$ for a typical UE. For a fixed UAV intensity $\lambda_{x}$, the ergodic rate monotonically increases with $\alpha$. This observation implies that the strength of unwanted interference decreases faster than the strength of the desired signal as $\alpha$ increases, thereby improving the ergodic rate. On the other hand, for a fixed $\alpha$, the ergodic rate decreases with $\lambda_{x}$ since larger $\lambda_{x}$ means more substantial interference. For either case mentioned above, the numerical results computed as per \eqref{Eq: ADR} in Theorem~\ref{Thm: typical} agree with the simulation ones, which corroborates the effectiveness of our performance analysis.
	
	\subsection{Comparison with the Conventional CoMP Schemes}
	\begin{figure}[!t]
		\centering
		\includegraphics[width=0.9\linewidth]{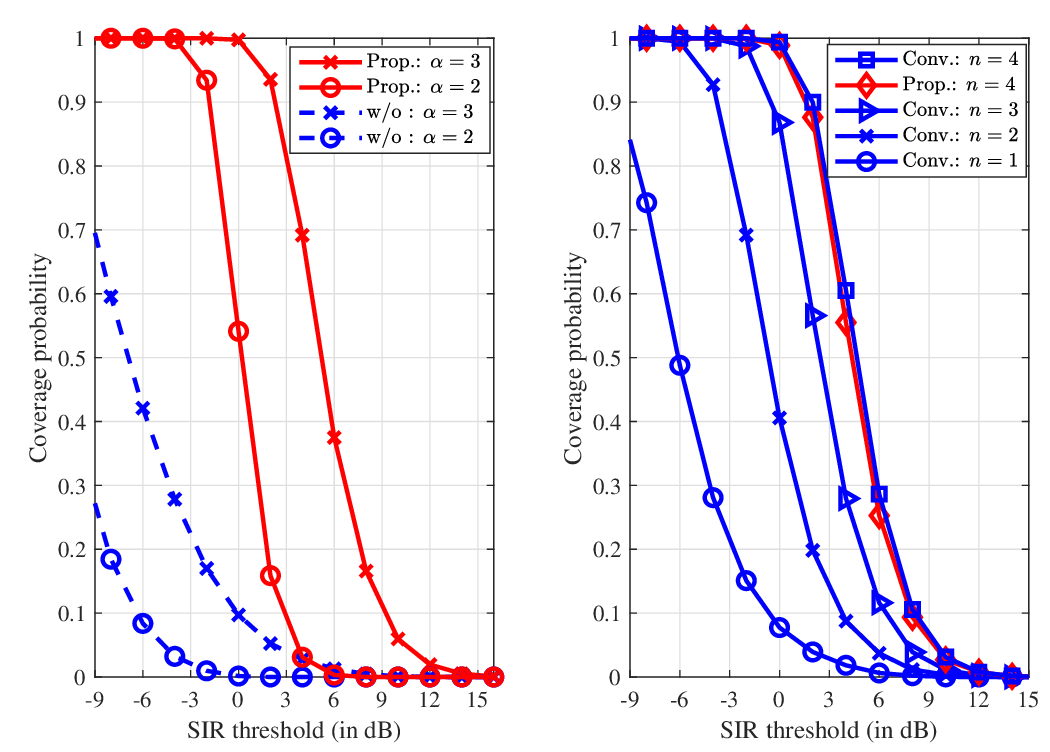} \vspace{-5pt}
		\caption{The coverage probabilities for a typical UE simulated under the proposed CoMP scheme and (left) the same scheme without CoMP, as well as (right) the conventional CoMP schemes with $n = \{1, 2, 3, 4\}$, with $\lambda_{x} = 16$ UAVs/\si{\km}$^2$.}
		\label{Fig: performance2}
	\end{figure}

	Figure~\ref{Fig: performance2} depicts the coverage probabilities for a typical UE under the proposed scheme and the conventional CoMP scheme with the nearest neighbor criterion (i.e., the fixed-number BS cooperation scheme in \cite{9644611}), with legends ``Prop.'' and ``Conv.'', respectively. For comparison purposes, the cases without CoMP for the proposed scheme (i.e., only the nearest UAV is chosen) are also considered, with the legend ``w/o''. The left panel of Fig.~\ref{Fig: performance2} depicts the coverage probability of a typical UE versus the SIR threshold, with the path loss exponents $\alpha = 2, \, 3$. It is seen that the coverage probability of our proposed scheme is much higher than that without CoMP transmission, even at a relatively high value of $\alpha = 3$. This manifests that CoMP transmission is highly beneficial. Moreover, comparing the curve with the legend ``w/o: $\alpha = 3$'' in the left panel of Fig.~\ref{Fig: performance2} to that with the legend ``Conv: $n = 1$'' in the right panel, their difference is caused by that the ``nearest'' UAV in the geometry sense is not precisely the nearest one in the sense of the Rx signal strength.
	
	The right panel of Fig.~\ref{Fig: performance2} compares the proposed scheme with the conventional CoMP scheme with $n = 1, 2, 3, 4$ dynamically collaborative UAVs. It is observed that the coverage probability increases with more CoMP UAVs, but the cooperative gain becomes smaller and smaller as there are more and more collaborative UAVs. When four collaborative UAVs are employed, the proposed scheme performs almost the same as the conventional CoMP scheme. This observation is in agreement with that reported in \cite[Fig. 18]{Li2020A2A}. In real-world applications, the conventional scheme needs exhaustive searching to find the CoMP set of UAVs, and no analytical performance evaluation is available. In contrast, the proposed Poisson-Delaunay simplex-based model avoids search overhead and allows analytical performance evaluation. 
	
	\section{Conclusions} \label{sec: conclusion}
	In this paper, a novel UAV-enabled air-to-ground network model was proposed. In particular, the CoMP transmission technique was employed to enhance the communication quality, associated with the idea of Delaunay tetrahedralization, whereby four UAVs can jointly serve each UE, thus providing reliable and high-throughput connectivity. Then, two movement control strategies were developed, including the UAV formation control for static UEs and UAV swarm tracking for mobile UEs. Afterward, analytical expressions regarding the coverage probability and ergodic rate were derived for a typical UE. Finally, simulation and numerical results collaborated that CoMP transmission brings significant cooperative gain compared to the transmission without CoMP, and the proposed scheme performs similarly to the conventional CoMP scheme with four collaborative UAVs while avoiding search overhead and facilitating performance evaluation, thanks to its mathematical tractable geometric model.  
	
\appendices
	
	\section{The proof of Theorem~\ref{Thm: formation}} \label{Appendix: C1}
	Construct the following Lyapunov functional candidate 
	\begin{equation}\label{Eq: c1-Lyapunov}
		V (t) = {\bm \epsilon}^{\text T} (t) {\bm R} \, {\bm \epsilon} (t),
	\end{equation}
	where ${\bm R} = \left[\begin{smallmatrix} c \, {\bm Q} & {\bm P} \\ {\bm P} & {\bm P} \end{smallmatrix}\right] \otimes \bm{I} _{m}$. Here, the matrix ${\bm P} = \text{Diag} \left\{p_{1}, \cdots, p_{n}\right\}$, with $p_{i} = 1 / q_{i}$ for all $i = 1, \cdots, n$, and ${\bm q} = \left[q_{1}, \cdots, q_{n}\right]^{\text T} = \left(\bm{L}  + \bm{B} \right)^{- 1} {\mathbbm 1}_{n}$. Based on \cite[Lem. 1]{Zhang2012}, the matrix ${\bm Q} = {\bm P} \left(\bm{L}  + \bm{B} \right) + \left(\bm{L}  + \bm{B} \right)^{\text T} {\bm P}$ is positive definite. Then by \cite[p.128]{Horn1991}, we obtain ${\bm R}$ is positive definite when $c > \lambda_{\max} ({\bm P}) / \lambda_{\min} ({\bm Q})$.
	
	Differentiating \eqref{Eq: c1-Lyapunov} with respect to $t$ yields~\eqref{Eq: proof-c1-xv}, shown on the bottom of this page. 
	\begin{figure*}[!b]
		\normalsize
		\setcounter{mytempeqncnt}{\value{equation}}
		\setcounter{equation}{46}
		\hrulefill
		\begin{align}
			\dot{V} (t)
			& = 2 {\bm \epsilon}^{\text T} (t) {\bm R}
			\left[\begin{smallmatrix}
				\bm{0} 							     	& \bm{I} _{n} \\
				- c \, \left(\bm{L}  + \bm{B} \right)	& - c \, \left(\bm{L}  + \bm{B} \right)
			\end{smallmatrix}\right]
			\otimes \bm{I} _{m} {\bm \epsilon} (t) + 2 {\bm \epsilon}^{\text T} (t) {\bm R} \otimes \bm{I} _{m}
			\left[\begin{smallmatrix}
				\bm{0} \\ {\bm F} \left(t, \bm{x} (t), \bm{v} (t)\right) - \bm{1}_{n} \otimes {\bm f} \left(t, \bm{x}_{0} (t), \bm{v}_{0} (t)\right)
			\end{smallmatrix}\right]
			\nonumber \\
			& = {\bm \epsilon}^{\text T} (t)
			\left[\begin{smallmatrix}
				- c \, {\bm Q}	& \bm{0} \\
				{\bm 0}			& 2 {\bm P} - c \, {\bm Q}
			\end{smallmatrix}\right]
			\otimes \bm{I} _{m} {\bm \epsilon} (t) + 2 \left[{\bm \xi}_{\bm{x}}^{\text T} (t) + {\bm \xi}_{\bm{v}}^{\text T} (t)\right] \left({\bm P} \otimes \bm{I} _{m}\right) \left[{\bm F} \left(t, \bm{x} (t), \bm{v} (t)\right) - \bm{1}_{n} \otimes {\bm f} \left(t, \bm{x}_{0} (t), \bm{v}_{0} (t)\right)\right].
			\label{Eq: proof-c1-xv} 
		\end{align}
		\setcounter{equation}{\value{mytempeqncnt}}
	\end{figure*}
    Based on the Assumption~\ref{Asump: Lipschitz}, we obtain
	\begin{align}
		\setcounter{equation}{47}
		& 2 {\bm \xi}_{\bm{x}}^{\text T} (t) \left({\bm P} \otimes \bm{I} _{m}\right) \left[{\bm F} \left(t, \bm{x} (t), \bm{v} (t)\right) - \bm{1}_{n} \otimes {\bm f} \left(t, \bm{x}_{0} (t), \bm{v}_{0} (t)\right)\right]
		\nonumber \\
		& \quad = 2 \sum_{i = 1}^{n} p_{i} {\bm \xi}_{\bm{x}, i}^{\text T} (t) \left[{\bm f} \left(t, \bm{x}_{i} (t), \bm{v}_{i} (t)\right) - {\bm f} \left(t, \bm{x}_{0} (t), \bm{v}_{0} (t)\right)\right]
		\nonumber \\
		& \quad \leq 2 \sum_{i = 1}^{n} p_{i} \left\|{\bm \xi}_{\bm{x}, i} (t)\right\| \left(\rho_{1} \left\|\bm{\xi}_{\bm{x}, i} (t)\right\| + \rho_{2} \left\|\bm{\xi}_{\bm{v},i} (t)\right\|\right)
		\nonumber \\
		& \quad \leq \left(2 \rho_{1} + \rho_{2}\right) \sum_{i = 1}^{n} p_{i} \left\|{\bm \xi}_{\bm{x}, i} (t)\right\|^{2} + \rho_{2} \sum_{i = 1}^{n} p_{i} \left\|{\bm \xi}_{\bm{v}, i} (t)\right\|^{2}.
		\label{Eq: proof-c1-x}
	\end{align}
	Similarly, we have~\eqref{Eq: proof-c1-v}, shown at the bottom of this page. 
	
	\begin{figure*}[!b]
		\normalsize
		\setcounter{mytempeqncnt}{\value{equation}}
		\setcounter{equation}{48}
		\hrulefill
		\begin{align}
			& 2 \bm{\xi}_{\bm{v}}^{\text T} (t) \left(\bm{P} \otimes \bm{I}_{m}\right) \left[\bm{F} \left(t, \bm{x} (t), \bm{v} (t)\right) - \bm{1}_{n} \otimes \bm{f} \left(t, \bm{x}_{0} (t), \bm{v}_{0} (t)\right)\right]
			\leq \rho_{1} \sum_{i = 1}^{n} p_{i} \left\|\bm{\xi}_{\bm{x}, i} (t)\right\|^{2} + \left(\rho_{1} + 2 \rho_{2}\right) \sum_{i = 1}^{n} p_{i} \left\|\bm{\xi}_{\bm{v}, i} (t)\right\|^{2}.
			\label{Eq: proof-c1-v}
		\end{align}
		\setcounter{equation}{\value{mytempeqncnt}}
	\end{figure*}
	
	Combine \eqref{Eq: proof-c1-xv} with \eqref{Eq: proof-c1-x}-\eqref{Eq: proof-c1-v}, we attain
	\begin{align}
		\setcounter{equation}{49}
		\dot{V} (t)
		& \leq {\bm \epsilon}^{\text T} (t)
		\left[\begin{smallmatrix}
			- c \, {\bm Q}	& \bm{0} \\
			{\bm 0}			& 2 {\bm P} - c {\bm Q}
		\end{smallmatrix}\right]
		\otimes \bm{I} _{m} {\bm \epsilon} (t) \nonumber\\
		& \quad + {\bm \epsilon}^{\text T} (t)
		\left[\begin{smallmatrix}
			\left(3 \rho_{1} + \rho_{2}\right) {\bm P}	& \bm{0} \\
			{\bm 0}										& \left(\rho_{1} + 3 \rho_{2}\right) {\bm P}
		\end{smallmatrix}\right]
		\otimes \bm{I} _{m} {\bm \epsilon} (t) \nonumber \\
		& < {\bm \epsilon}^{\text T} (t) {\bm S} {\bm \epsilon} (t),
		\label{Eq: proof-c2-derivatives}
	\end{align}
	where ${\bm S} = \left[\begin{smallmatrix} 	- c \, {\bm Q} + d {\bm P} & \bm{0} \\ {\bm 0} & - c \, {\bm Q} + d {\bm P} \end{smallmatrix}\right] \otimes \bm{I} _{m}$, with $d = \max \left\{3 \rho_{1} + \rho_{2}, \rho_{1} + 3 \rho_{2} + 2\right\}$.
	
	By virtue of \eqref{Eq: proof-c2-derivatives}, it is straightforward that $\dot{V} (t) \leq 0$ when $c > d \, \lambda_{\max} (\bm{P}) / \lambda_{\min} (\bm {Q})$, and $\dot{V} (t) = 0$ if and only if $\bm{\xi}_{\bm{x}} (t) = \bm{0}$ and $\bm{\xi}_{\bm{v}} (t) = \bm{0}$. This completes the proof.
	
	\section{The proof of Theorem~\ref{Thm: tracking}} \label{Appendix: C2}
	To prove Theorem~\ref{Thm: tracking} is equivalent to prove the following two lemmas. 	
	
	\begin{lemma} \label{Lem: proof-c2-1}
		Any solution of the error dynamics \eqref{Eq: C2-error} with an initial value being out of $\mathcal{Q}_{0}$ can reach $\mathcal{Q}_{0}$ if the time $t_{k}$ is sufficiently large.
	\end{lemma}
	
	\begin{IEEEproof}
		The proof is by contradiction. Assume there exists a solution $\bm{\epsilon_{g}}(t)$ of \eqref{Eq: C2-error} with an initial value out of $\mathcal{Q}_{0}$, i.e., $\bm{\epsilon_{g}}(t_{0}) \in \mathcal{Q}_{0}^{c}$, and it cannot reach $\mathcal{Q}_{0}$ as the time $t_{k} \to \infty$. 
		
		Construct the Lyapunov functional candidate as
		\begin{equation} \label{Eq: c2-Lyapunov}
			V \left(t\right) = \left[\bm{\epsilon}_{g}^{F}(t)\right]^{\rm T} \bm{\epsilon}_{g}^{F} (t) + \left[\bm{\epsilon}_{g}^{L}(t)\right]^{\rm T} \bm{\epsilon}_{g}^{L} (t).
		\end{equation}
		For any $t \in (t_{k - 1}, t_{k}]$, the right and upper Dini’s derivative of $V (t)$ along the trajectory is~\eqref{Eq: proof-c2-dini}. 
		\begin{figure*}[!b]
			\normalsize
			\setcounter{mytempeqncnt}{\value{equation}}
			\setcounter{equation}{51}
			\hrulefill
			\begin{align}
				\mathcal{D}^{+} V (t)
				& = \left[\dot{\bm{\epsilon}}_{g}^{F}(t)\right]^{\rm T} \bm{\epsilon}_{g}^{F} (t) + \left[\bm{\epsilon}_{g}^{F}(t)\right]^{\rm T} \dot{\bm{\epsilon}}_{g}^{F} (t) + \left[\dot{\bm{\epsilon}}_{g}^{L}(t)\right]^{\rm T} \bm{\epsilon}_{g}^{L} (t) + \left[\bm{\epsilon}_{g}^{L}(t)\right]^{\rm T} \dot{\bm{\epsilon}}_{g}^{L} (t)
				\nonumber \\
				& = \left[\bm{\epsilon}_{g}^{F}(t)\right]^{\rm T} \left[\left({\bm G}^{F}\right)^{\rm T} + {\bm G}^{F}\right] \bm{\epsilon}_{g}^{F} (t) + 2
				\left[\begin{smallmatrix} 
					\bm{0} \\ \bm{F} (t) - {\mathbbm 1_{n}} \otimes {\bm g} (t)
				\end{smallmatrix}\right]^{\rm T} \bm{\epsilon}_{g}^{F} (t)
				\nonumber \\
				& \quad + \left[\bm{\epsilon}_{g}^{L}(t)\right]^{\rm T} \left[\left({\bm G}^{L}\right)^{\rm T} + {\bm G}^{L}\right] \bm{\epsilon}_{g}^{L} (t) + 2
				\left[\begin{smallmatrix}
					\bm{0} \\ \bm{f}_{0} (t) - {\mathbbm 1_{n}} \otimes {\bm g} (t)
				\end{smallmatrix}\right]^{\rm T} \bm{\epsilon}_{g}^{L} (t)
				\nonumber \\
				& \leq \underbrace{\left\{\max \left\{\left[\left({\bm G}^{F}\right)^{\rm T} + {\bm G}^{F}\right], 1\right\} + 2 \left(\max \left\{\left\|{\bm F}(t)\right\|_{\infty}, \left\|{\bm f}_{0} (t)\right\|_{\infty} \right\}\right) + 2 \left\|{\bm g} (t)\right\|_{\infty}\right\}}_{\eta} V(t).	
				\label{Eq: proof-c2-dini}
			\end{align}
		    
		    \begin{align}
		    	V (t_{k}^{+})
		    	& = \left[\bm{\epsilon}_{g}^{F} (t_{k}) + \Delta \bm{\epsilon}_{g}^{F} (t_{k}) + \bm{\epsilon}_{g}^{L} (t) +  \Delta \bm{\epsilon}_{g}^{L} (t) \right]^{\rm T} \left[\bm{\epsilon}_{g}^{F} (t_{k}) + \Delta \bm{\epsilon}_{g}^{F} (t_{k}) + \bm{\epsilon}_{g}^{L} (t) +  \Delta \bm{\epsilon}_{g}^{L} (t) \right]
		    	\nonumber \\
		    	& = \left[\bm{\epsilon}_{g}^{F}(t)\right]^{\rm T} \left(\bm{I}_{m n}  + {\bm H}_{k}^{F}\right)^{\rm T} \left(\bm{I}_{m n}  + {\bm H}_{k}^{F}\right) \bm{\epsilon}_{g}^{F}(t) + \left[\bm{\epsilon}_{g}^{L}(t)\right]^{\rm T} \left(\bm{I}_{2 m}  + {\bm H}_{k}^{L}\right)^{\rm T} \left(\bm{I}_{2 m} + {\bm H}_{k}^{L} \right) \bm{\epsilon}_{g}^{L} (t)
		    	\nonumber \\
		    	& \leq \underbrace{\max \left\{\lambda_{\max} \left(\bm{I}_{m n}  + {\bm H}_{k}^{F}\right), \lambda_{\max} \left(\bm{I}_{2 m} + {\bm H}_{k}^{L} \right)\right\}}_{\beta_{k}} V (t_{k}).
		    	\label{Eq: proof-c2-tkplus}
		    \end{align}
			\setcounter{equation}{\value{mytempeqncnt}}
		\end{figure*}
		Moreover, for any $t \in (t_{k - 1}, t_{k}]$, we have $V (t) \leq V (t_{k - 1}^{+}) \exp\{\eta (t - t_{k})\}$. On the other hand, when the system adopts the impulse control~\eqref{Eq: C2-l}, by \eqref{Eq: C2-impulse}, we have~\eqref{Eq: proof-c2-tkplus}, shown on the bottom of the next page.

		Let $k = 1$, i.e., $t \in (t_{0}, t_{1}]$, we obtain $	V(t_{1}^{+}) \leq \beta_{1} V (t_{1}) \leq \beta_{1} V (t_{0}) \exp \left\{\eta (t_{1} - t_{0})\right\}$, and for $t \in (t_{1}, t_{2}]$, a similar result is given by $V (t_{2}^{+}) \leq \beta_{2} V (t_{0}) \leq \beta_{1} \beta_{2} V (t_{0}) \exp \left\{\eta (t_{2} - t_{0})\right\}$. In general, for $t \in (t_{k - 1}, t_{k}]$, we attain
		\begin{equation} \label{Eq-54}
			\setcounter{equation}{54}
			V (t) \leq \prod_{k = 1}^{n} \beta_{k} V (t_{0}) \exp \left\{\eta (t - t_{0})\right\}.
		\end{equation}
		By using \cite[Thm. 3.1.4]{Yang2001} together with $\eta (t_{k} - t_{k - 1}) + \ln(\rho \beta_{k})< 0$, for $t \in (t_{k - 1}, t_{k}]$, \eqref{Eq-54} is further bounded by 
		\begin{align}  \label{Eq-55}
			V(t)  & \leq \prod_{k = 1}^{n} V (t_{0}) \exp \left(\eta (t - t_{k})\right) \exp \left\{\eta (t_{k} - t_{0})\right\} \nonumber\\
			& \leq  \frac{1}{\rho^{k}} V (t_{0}) \exp \left\{\eta (t - t_{k})\right\}.
		\end{align}
		Then, with $\rho > 1$, \eqref{Eq-55} gives 
		\begin{equation}
			\lim_{k \to \infty} V (t) = 0, \quad \forall t \in (t_{k}, t_{k + 1}],
		\end{equation}
		which means $\lim_{k \to \infty} \bm{\epsilon}_{g}^{F} (t) = \bm{0}$, and $\lim_{k \to \infty} \bm{\epsilon}_{g}^{L} (t) = \bm{0}$ for $t \in (t_{k - 1}, t_{k}]$. This clearly contradicts the original assumption. Therefore, a solution does not exist, as asserted above.
	\end{IEEEproof}
	
	\begin{lemma}\label{Lem: proof-c2-2}
		There exists a state admissible set $\mathcal{Q} \supseteq \mathcal{Q}_{0}$ that for any solution of \eqref{Eq: C2-error} in $\mathcal{Q}_{0}$ cannot exceed it.
	\end{lemma}
	
	\begin{IEEEproof}
		Introducing an auxiliary system of the UAV swarm \eqref{Eq: C2-general} as the special case that the leader's planar projection is the same as the UE's location. For $t \in (t_{k - 1}, t_{k}]$, we have
		\begin{subequations}\label{Eq: proof-c2-aiderror}
			\begin{align}
				\mathcal{D}^{+} \bm{e} (t) & =
				\bm{M} {\bm e} (t) +
				\left[\begin{smallmatrix}
					\bm{0} \\
					{\bm F (t)} - \mathbbm{1}_{n} \otimes {\bm f}_{0} (t)
				\end{smallmatrix}\right], \ t \neq t_{k};
				\label{Eq: proof-c2-aiderror-a} \\
				\Delta \bm{e}_{i} (t_{k}) & =
				\begin{cases}
					\left[\begin{smallmatrix}
						{\Delta \bm{x}}_{i} (t_{k}) 	& \bm{0} \\
						\bm{0} 					& {\Delta \bm{v}}_{i} (t_{k})
					\end{smallmatrix}\right] \otimes \bm{I} _{m},  \\
					\qquad \qquad \quad \text{if a UAV outside of} \ \mathcal{Q}_{0}; \\
					\bm{0}, \qquad \qquad \text{otherwise},
				\end{cases}
				\label{Eq: proof-c2-aiderror-b}
			\end{align}
		\end{subequations}
		with $\bm{M} = \left[\begin{smallmatrix} \bm{0}  & \bm{I}  \\ - c \, \bm{L}  & - c \, \bm{L} \end{smallmatrix}\right] \otimes \bm{I} _{m}$, and ${\Delta \bm{x}}_{i} (t_{k}) $ and ${\Delta \bm{v}}_{i} (t_{k})$ are the same as \eqref{Eq: C2-impulse} for all $i$.
		
		For the case that there exists a solution $\bm{e} (t)$ is out of $\mathcal{Q}_{0}$ at time $t$, we choose $V_{1}(t) = {\bm{e}}^{\text{T}} (t) {\bm{e}} (t)$ as the Lyapunov function. Using a similar trick as in Lemma~\ref{Lem: proof-c2-1}, we obtain $\mathcal{D}^{+} V_{1} (t) \leq \eta^{\prime} V_{1} (t)$, with $\eta^{\prime} = \lambda_{\max}(\bm{M} + \bm{M}^{\rm T}) + 2 \|\bm{F} (t)\|_{\infty} + 2 \|\bm{f}_{0} (t)\|_{\infty}$. This means that the solution ${\bm{e}} (t + \Delta t)$ eventually reach $\mathcal{Q}_{0}$ as $\Delta t$ is sufficiently large. The dotted portion of the ``Trajectory-A'' in Fig.~\ref{Fig: stateset} illustrates this process. On the other hand, considering the solution $\bm{e} (t)$ is in $\mathcal{Q}_{0}$, and using a similar method as in Appendix~\ref{Appendix: C1}, it is straightforward to infer that the solutions of \eqref{Eq: proof-c2-aiderror} are in $\mathcal{Q}_{0}$, and they are asymptotically stable (at origin). That is, there exists a time $t^{\prime}$, for all $t > t^{\prime}$, the solution $\bm{e} (t)$ of \eqref{Eq: proof-c2-aiderror} in $\mathcal{Q}_{0}$ satisfies $\|\bm{e} (t)\| < \varepsilon$ for any $\varepsilon \in \mathbb{R}^{+}$.
	\end{IEEEproof}
	
	Through the proof of the above two lemmas, the multi-agent tracking system keeps the required formation and remains in the state admissible set $\mathcal{Q}$ corresponding to $\mathcal{Q}_{0}$ when $t_{k}$ is sufficiently large if there exists $\rho > 1$ such that the condition $\eta (t_{k} - t_{k - 1}) + \ln(\rho \beta_{k})< 0$ (i.e., Eq.~\eqref{Eq: C2-thm1} in Theorem~\ref{Thm: tracking}) establish for all $k$. Moreover, according to the theory of kinematics, the radius of state admissible set $\mathcal{Q}$ can be expressed as~\eqref{Eq: C2-thm2}, which is obtained by calculating the variations of the UAVs' velocities and the distances between each UAV and the mobile UE.

	\section{The proof of Lemma~\ref{Lem: T-pdf}} \label{Appendix: A1}
	Using the central limit theorem for causal functions \cite{Papoulis1962}, the PDF of $T \triangleq \sum_{i} d_{i}^{-\alpha / 2} \|{h}_{i}\|$ can be well approximated by the Gamma distribution \eqref{Eq: T-apdf} with the parameters
	\begin{equation}\label{Eq: Gamma-paras}
		\nu = \frac{\mathbb{E}^{2}[T]} {{\text{Var}}[T]}, {\text{ and }}  \theta = \frac{{\text{Var}}[T]} {\mathbb{E}[T]}.
	\end{equation}
	
	Since $d_{i}$ and $\|{\bm h}_{i}\|$ are independent of each other, the expectation of $T$ can be computed as
	\begin{align*}
		\mathbb{E}[T]
		& = \mathbb{E}\left[\|{h}_{i}\| \right] \left(\sum_{i = 1}^{4} \mathbb{E}\left[d_{i}^{- \frac{\alpha}{2}} \right]\right),
	\end{align*}
	and the variance of $T$ is computed as
	\begin{align*}
		{\text{Var}}[T]
		& = \mathbb{E}\left[\|{h}_{i}\|^{2} \right] \left(\sum_{i = 1}^{4} \mathbb{E}\left[d_{i}^{- \alpha} \right]  \right. \nonumber \\
		& \quad {}+ \left. \sum_{i \neq j} \mathbb{E}\left[d_{i}^{- \frac{\alpha}{2}} \right] \mathbb{E}\left[d_{j}^{- \frac{\alpha}{2}} \right]  \right) - \left(\mathbb{E}[T] \right)^{2}.
	\end{align*}
	Recalling \eqref{Eq: nakagami}, we have
	\begin{equation*}
		\mathbb{E}\left[\|{h}_{i}\| \right] = \frac{\Gamma(m + \frac{1}{2})}{\Gamma(m)} \left(\frac{\Omega}{m}\right)^{\frac{1}{2}}, {\text { and }} \mathbb{E}\left[\|{h}_{i}\|^{2} \right] = \Omega.
	\end{equation*}
	Finally, by \eqref{Eq: dn-pdf}, the expectation of $d_{i}^{- n}$ can be readily computed by \eqref{Eq: T-para3}. Consequently, the parameters $\nu$ and $\theta$ in \eqref{Eq: T-apdf} can be calculated by \eqref{Eq: T-para1} and \eqref{Eq: T-para2}, respectively.

	\bibliographystyle{IEEEtran}
	\bibliography{reference}	
	
	\begin{IEEEbiography}
		[{\includegraphics[width=1in, height=1.25in, clip, keepaspectratio]{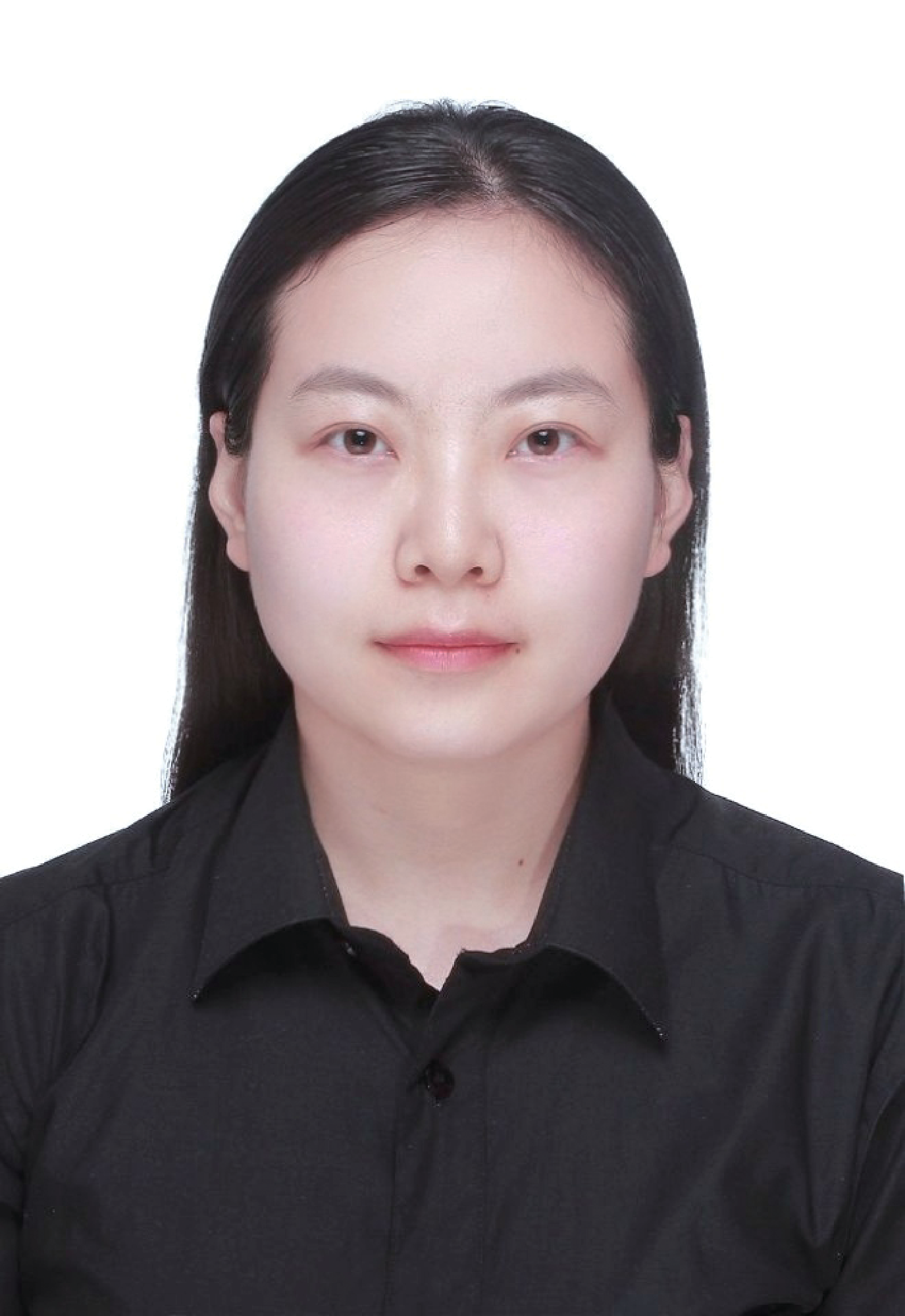}}]{Xiao Fan} received the B.S. and the M.S. degree in mathematics from the Shanxi Datong University, Datong, China, in 2015, and the Guilin University of Electronic Technology, Guilin, China, in 2018, respectively. She is currently pursuing a Ph.D. degree in information and communication engineering at Sun Yat-sen University, Guangzhou, China. Her research interests include cooperative UAV communications and nonlinear systems.
	\end{IEEEbiography}
	
	\begin{IEEEbiography}
		[{\includegraphics[width=1in, height=1.25in, clip, keepaspectratio]{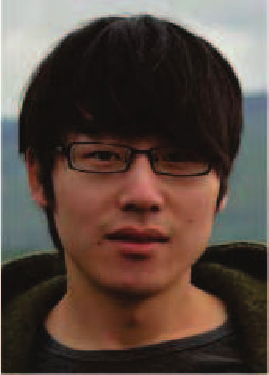}}]{Peiran Wu } (Member, IEEE) received a Ph.D. degree in electrical and computer engineering from The University of British Columbia (UBC), Vancouver, Canada, in 2015.
		
		From October 2015 to December 2016, he was a Post-Doctoral Fellow at UBC. In the Summer of 2014, he was a Visiting Scholar with the Institute for Digital Communications, Friedrich-Alexander-University Erlangen-Nuremberg (FAU), Erlangen, Germany. Since February 2017, he has been with Sun Yat-sen University, Guangzhou, China, where he is currently an Associate Professor. Since 2019, he has been an Adjunct Associate Professor with the Southern Marine Science and Engineering Guangdong Laboratory, Zhuhai, China. His research interests include mobile edge computing, wireless power transfer, and energy-efficient wireless communications.
				
		Dr. Wu was a recipient of the Fourth-Year Fellowship in 2010, the C. L. Wang Memorial Fellowship in 2011, the Graduate Support Initiative (GSI) Award from UBC in 2014, the German Academic Exchange Service (DAAD) Scholarship in 2014, and the Chinese Government Award for Outstanding Self-Financed Students Abroad in 2014.
	\end{IEEEbiography}
	
	\begin{IEEEbiography}
		[{\includegraphics[width=1in, height=1.25in, clip, keepaspectratio]{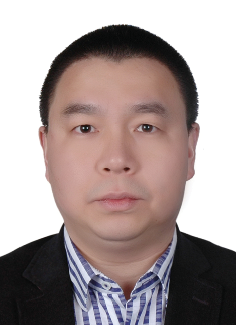}}]{Minghua Xia} (Senior Member, IEEE) received the Ph.D. degree in Telecommunications and Information Systems from Sun Yat-sen University, Guangzhou, China, in 2007.
		
		From 2007 to 2009, he was with the Electronics and Telecommunications Research Institute (ETRI) of South Korea, Beijing R\&D Center, Beijing, China, where he worked as a member and then as a senior member of the engineering staff. From 2010 to 2014, he was in sequence with The University of Hong Kong, Hong Kong, China; King Abdullah University of Science and Technology, Jeddah, Saudi Arabia; and the Institut National de la Recherche Scientifique (INRS), University of Quebec, Montreal, Canada, as a Postdoctoral Fellow. Since 2015, he has been a Professor at Sun Yat-sen University. Since 2019, he has also been an Adjunct Professor with the Southern Marine Science and Engineering Guangdong Laboratory (Zhuhai). His research interests are in the general areas of wireless communications and signal processing.
	\end{IEEEbiography}

\end{document}